\documentstyle[eqsecnum,epsf,aps,amsmath,amssymb,amsbsy,multicol]{revtex}

\begin{document}
%%\draft

\title{Phase Separation of Rigid-Rod Suspensions in Shear Flow}
\author{Peter~D. Olmsted$^{1}$ and C.-Y.~David~Lu$^{2}$}
\address{$^1$Department of Physics, University of Leeds, Leeds LS2
  9JT, UK {\bf\tt (p.d.olmsted@leeds.ac.uk)}
  and $^2$Department  of Physics and   Center of Complex
  Systems,  National Central University, Chung-li, 320 Taiwan {\bf\tt
    (dlu@joule.phy.ncu.edu.tw)}} \date{\today} \maketitle
\def\thefootnote{\fnsymbol{footnote}}

%%%%%%%%%%%%%%%%%%%%%%%%%%%%%%%%%%%%%%%%%%%%%%%%%%%%%%%%%%%%%%%%%%%%%%%%%%%%%
\begin{abstract} 
  We analyze the behavior of a suspension of rigid rod-like particles
  in shear flow using a modified version of the Doi model, and
  construct diagrams for phase coexistence under conditions of
  constant imposed stress and constant imposed strain rate, among
  paranematic, flow-aligning nematic, and log-rolling nematic states.
  We calculate the effective constitutive relations that would be
  measured through the regime of phase separation into shear bands. We
  calculate phase coexistence by examining the stability of
  interfacial steady states and find a wide range of possible
  ``phase'' behaviors.
\end{abstract}
  %%, (Transitions in liquid crystals), 
%%%%%%%%%%%%%%%%%%%%%%%%%%%%%%%%%%%%%%%%%%%%%%%%%%%%%%%%%%%%%%%%%%%%%%%%%%%%%
\begin{multicols}{2}
%%%%%%%%%%%%%%%%%%%%%%%%%%%%%%%%%%%%%%%%%%%%%%%%%%%%%%%%%%%%%%%%%%%%%%%%%%%%%
%%%%%%%%%%%%%%%%%%%%%%%%%%%%%%%%%%%%%%%%%%%%%%%%%%%%%%%%%%%%%%%%%%%%%%%%%%%%%
\section{Introduction}
%%%%%%%%%%%%%%%%%%%%%%%%%%%%%%%%%%%%%%%%%%%%%%%%%%%%%%%%%%%%%%%%%%%%%%%%%%%%%

Shear flow has profound effects on complex fluids. It can perturb
equilibrium phase transitions, such as the isotropic-to-nematic (I-N)
liquid crystalline transition in wormlike micelles
\cite{berret94a,berret94b,BPD97}, thermotropic melts
\cite{hess76,olmsted90,olmsted92,mather97}, or rigid-rod suspensions
\cite{see90,olmstedlu97}; the nematic-smectic transition in
thermotropic liquid crystals \cite{safinya91}; and the
isotropic-to-lamellar transition \cite{catesmilner89} in surfactant
systems.  Shear can also induce structures, such as the well-known
multi-lamellar vesicles (onions) in surfactant systems
\cite{roux93,diat93,diat95}, that exist only as metastable equilibrium
phases. Another well-known effect is the transition between
orientations of diblock copolymer lamellae in either the steady shear
flow \cite{fredrickson94,goulian95}, or the oscillatory shear flow
\cite{koppi92,winey93,kannan94}, as a function of shear rate or
frequency, and temperature.

%%%%%%%%%%%%%%%%%%%%%%%%%%%%%%%%%%%%%%%%%%%%%%%%%%%%%%%%%%%%%%%%%%%
 \begin{figure}[p]
 \par\columnwidth20.5pc
 \hsize\columnwidth\global\linewidth\columnwidth
 \displaywidth\columnwidth
 \epsfxsize=3.5truein
 \centerline{\epsfbox[30 12 700 440]{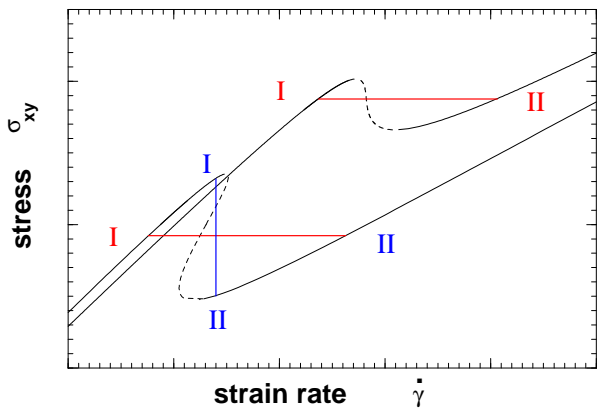}}
 \caption{ Stress--strain-rate curves for the Doi
   model with different excluded volume parameters $u$ (taken from
   Fig.~\protect{\ref{fig:constit}} below). The dashed line segments
   are unstable (unphysical) steady states. The straight lines indicate
   possible coexistence between states $I$ and $II$ under conditions of
   common stress (horizontal lines) or strain rate (vertical
   line).}
 \label{fig:both}
 \end{figure}
%%%%%%%%%%%%%%%%%%%%%%%%%%%%%%%%%%%%%%%%%%%%%%%%%%%%%%%%%%%%%%%%%%%
A related phenomenon is dynamic instability in non-Newtonian fluids
whose theoretical {\sl homogeneous\/} stress--strain-rate constitutive
relations exhibit multi-valued behavior, as in theories of polymer
melts \cite{doiedwards,catesmcleish93} and wormlike micelles
\cite{spenley93,cates90,rehage91}. Such models may describe, for
example, the spurt effect, whereby the flow rate of a fluid in a pipe
changes discontinuously as a function of applied pressure drop
\cite{mcleish86}.  A non-monotonic constitutive curve as in
Fig.~\ref{fig:both} typically has a segment (shown as a broken line)
where bulk flow is unstable. If a mean strain rate is imposed which
forces the system to lie on an unstable part of the constitutive
relation, a natural resolution for this instability is to break the
system into two regions, often called \textit{bands}, one on the high
strain rate branch and one on the low strain rate branch, to maintain
the overall applied strain rate. The most important unresolved
question about these banded flows is, what determines the stress at
which the system phase separates into bands?  Experiments on many
systems (reviewed in Sec.~IIA), particularly the wormlike micelle
surfactant systems, reveal that there is a well-defined and
reproducible selected stress in a wide class of systems

There have been many suggestions for determining the selected stress.
Some workers have assumed the existence of a non-equilibrium potential
and a variational principle \cite{Zuba96,porte97}. This possibility is
intriguing, although it remains unproven.  Early studies postulated a
jump at the top of the stable viscous branch (``top jumping'')
\cite{catesmcleish93,spenley93,schmitt95}, but experiments have shown
that this is not the case \cite{grand97}.  Recent studies have solved
the homogeneous flow equations in various geometries using
sophisticated hydrodynamic flow-solvers and found a selected stress
\cite{greco97,espanol96}.  However, evidence is growing \cite{history}
that these calculations have history-dependent stress selection (which
is in fact no selection) or introduce gradient terms due to the
discretization of the system.  A final method, which we follow here,
has been to incorporate (physically present) non-local contributions
to the stress
\cite{olmsted90,olmsted92,pearson94,spenley96,olmstedlu97,history,jsplanar,goveas99,dhont99},
and examine the equations of motion under steady banded flow
conditions.

Here we extend previous work \cite{olmstedlu97} and calculate phase
diagrams for rigid-rod suspensions in shear flow, solving for the
interfacial profile between phases and using its properties to
determine the coexistence stress. As Fig.~\ref{fig:both} indicates,
phase separation is possible at \textit{either} a specified stress
(horizontal tie lines) \textit{or} a specified strain rate (vertical
tie lines). Only recently has the latter possibility been speculated
upon \cite{schmitt95,olmstedlu97,porte97}, and found experimentally
\cite{Bonn+98}.  We explore this possibility explicitly for our model
system, which possesses, in addition to the high and low strain rate
(paranematic and nematic, respectively) branches shown in
Fig.~\ref{fig:both}, a second high strain rate branch in which the
rods stand up in the flow, parallel to the vorticity direction,
instead of lying in the shear plane \cite{bhave93}. We study
coexistence with this so-called `log-rolling' phase and find a rich
non-equilibrium phase diagram.

The summary of this paper is as follows. In Section~II we discuss the
general issues of shear banding and phase separation in flow, and
summarize the primary experimental evidence for this behavior. In
Section~III we present the modified Doi model \cite{doi81,doikuzuu83}
and in Section~IV we briefly discuss our algorithm for calculating the
phase diagram. The general aspects of the interface construction will
be discussed elsewhere \cite{jsplanar}.  We present the results for
common stress and strain rate phase separation in Section~V and~VI,
respectively, and discuss some of the implications for metastability
and experiments under controlled stress or controlled strain rate
conditions.  We finish in Section~VII with a discussion and summary.
While some of these results have been briefly summarized elsewhere
\cite{faraday}, the current paper is a complete and self-contained
discussion of the problem.

The reader interested in the phenomenology of phase diagrams for
sheared complex fluids rather than liquid crystals may safely skip
Section~III; the rest of the paper is general, and much of the
discussion applies to any system undergoing phase separation in shear
flow. There are, essentially, two steps to calculating phase behavior
in flow. One must derive the dynamical equations of motion for fluid
flow, composition, and the relevant structural order parameter(s),
which is quite difficult. Then, one must understand how to solve them
and interpret the results.  While the modified Doi model does not
exhaust all possible phase diagrams (in particular, a shear-thickening
model would be a nice complement), it has many universal features. One
extremely important concept is that density and field variables are
ill-defined in non-equilibrium systems: {\sl either\/} stress {\sl
  or\/} strain rate may act as a control parameter analogous to an
equilibrium field variable (\textit{e.g. }pressure, chemical
potential), corresponding to the different orientations of the
interface between coexisting phases.  Also, one can gain much
intuition from the underlying stress--strain-rate--composition
\textit{surface}, a fact which we feel has been underappreciated until
now.

%%%%%%%%%%%%%%%%%%%%%%%%%%%%%%%%%%%%%%%%%%%%%%%%%%%%%%%%%%%%%%%%%%%%%%%%%%%%
\section{Shear Banding}
%%%%%%%%%%%%%%%%%%%%%%%%%%%%%%%%%%%%%%%%%%%%%%%%%%%%%%%%%%%%%%%%%%%%%%%%%%%%
%%%%%%%%%%%%%%%%%%%%%%%%%%%%%%%%%%%%%%%%%%%%%%%%%%%%%%%%%%%%%%%%%%%%%%%%%%%%
\subsection{Experimental Evidence}
%%%%%%%%%%%%%%%%%%%%%%%%%%%%%%%%%%%%%%%%%%%%%%%%%%%%%%%%%%%%%%%%%%%%%%%%%%%%
Shear banding has been confirmed in many systems through direct
optical and NMR visualization, and deduced from rheological
measurements. The best-studied systems are surfactant solutions of
various kinds, including wormlike micelles and onion-lamellar phases.
Rehage and Hoffmann \cite{rehage91} measured a plateau in the
stress--strain-rate relation for wormlike micelles in shear flow.
This behavior has since been seen in a number of wormlike micellar
systems in various flow geometries, by the Montpellier
\cite{berret94a,berret94b,BPD97}, Strasbourg \cite{schmitt94}, Edinburgh
\cite{grand97}, and Massey groups
\cite{Call+96,MairCall96,BritCall97,MairCall97}. Berret \textit{et
  al.}  \cite{BPD97} visualized shear bands in the plateau region of
the stress--strain-rate curves using optical techniques, providing
proof of banding; and Callaghan \textit{et al.}
\cite{Call+96,MairCall96,BritCall97,MairCall97} used NMR to measure
the velocity profile in various geometries (including Couette,
cone-and-plate, and pipe geometries).

The transition in these cases is to a strongly-aligned, possibly
nematic, phase of wormlike micelles which has a lower viscosity than
the quiescent phase. It is not known how the length distribution
changes in flow, although this is certainly an important aspect of
these `living' systems \cite{Vand94}. Wormlike micellar system can
possess an equilibrium nematic phase, and in some cases the
shear-induced phase is obviously influenced by the proximity of an
underlying nematic phase transition
\cite{berret94a,berret94b,Roux+95,BPD97,Capp+97}.  However, many
wormlike micellar systems undergo banding at compositions much more
dilute than that for I-N coexistence, and it is probable that in these
cases flow instability is due to the non-linear rheology of these
systems, which is in many respects similar to that of the Doi-Edwards
model of polymer melts \cite{cates90}. Since there are at lease two
possible effects (a nematic phase transition and flow-instability of
the micellar constitutive relation) apparently leading to
flow-instability, these systems are quite rich.  It is tempting to
analyze the extent to which these systems display behavior analogous
to the kinetics of equilibrium phase separation, and groups have
recently begun to study the kinetics of non-equilibrium phase
separation \cite{berret94b,grand97,Berr97}.

Pine and co-workers have recently studied a wormlike surfactant system
at extreme dilutions and found, surprisingly, that for low enough
concentrations (but still above the overlap concentration) shear
induces a viscoelastic phase that they interpret as a gel
\cite{boltenhagen97a,boltenhagen97b,keller97}.  The origins and
structure of this gel are currently unknown. In controlled stress
experiments they observe shear banding and a `plateau' for stresses
higher than a certain stress, in which the strain rate
\textit{decreases} as shear induces the gel. Above the stress at which
the gel fills the sample cell the strain rate increases again to
complete a dramatic \textsf{S} curve. For controlled strain rate
experiments the system jumps, at a well defined strain rate, between
the gel and solution phases.

Another well-studied system is the onion lamellar surfactant phase,
originally studied by Roux, Diat, and Nallet
\cite{roux93,diat93,diat95}. These systems display a bewildering
variety of transitions between lamellar, aligned-lamellar, onion, and
onion crystal phases of various symmetries, as functions of applied
shear flow, temperature, and composition. As an example, one
particular system undergoes transitions, with increasing strain rate,
from disordered lamellae to onion, to onion-lamellae coexistence (in
which coexistence is inferred from a plateau in the
stress--strain-rate curves), to well-ordered lamellae \cite{diat93}.
Recently Bonn and co-workers \cite{Bonn+98} found shear-induced
transitions between different gel states of lamellar onion solutions
with shear bands (visualized by inserting tracer particles) oriented
with interface normals in the vorticity direction, indicating phase
separation at common strain rate instead of common stress, as we
clarify below. In this case the averaged stress--strain-rate
constitutive relation followed a sideways \textsf{S} curve under
controlled strain rate conditions.

Mather \textit{et al.} \cite{mather97} have recently studied a
thermotropic polymer liquid crystal using visual and rheological
measurements, and inferred a shear-induced nematic phase transition
and phase separation, the latter which they attribute to
polydispersity.

In summary, shear-banding has been seen in several systems, and in all
cases is associated with some flow-induced change in the fluid
microstructure.  Most systems are still poorly-understood
\cite{boltenhagen97a,diat95,Bonn+98} and, given the range of complexity,
it is certain that many qualitatively new phenomena remain to be
discovered.

%%%%%%%%%%%%%%%%%%%%%%%%%%%%%%%%%%%%%%%%%%%%%%%%%%%%%%%%%%%%%%%%%%%%%%%%%%%%
\subsection{Theoretical Issues}
%%%%%%%%%%%%%%%%%%%%%%%%%%%%%%%%%%%%%%%%%%%%%%%%%%%%%%%%%%%%%%%%%%%%%%%%%%%%
The crux of the problem from a theoretical point of view may be
appreciated from Fig.~\ref{fig:both}.  These stress--strain-rate
curves are somewhat reminiscent of pressure-density ($p-\rho$)
isotherms for a liquid-gas system. Curve segments with negative slope,
$\partial\sigma_{xy}/\partial\dot{\gamma}<0$, are unstable and cannot
describe a physical state of a bulk homogeneous system.  Analogously,
isotherms with negative slopes $\partial p/\partial\rho < 0$ have
negative bulk moduli and are unstable. The liquid-gas system resolves
this instability by phase-separating into regions of different
densities (according to the lever rule to maintain the average
density). Similarly, the banded flows seen in the experiments
described above appear to be a non-equilibrium phase separation into
regions of high and low strain-rate, maintaining the applied mean
strain rate.

In previous work \cite{olmsted90,olmsted92,olmstedlu97} we constructed
a `phase diagram' by pursuing an analogy between homogeneous stable
steady states and equilibrium phases.  As in equilibrium,
non-equilibrium `phases' may be separated, in field variable space, by
hyper-surfaces representing continuous ({\sl e.g.\/} critical
points/lines) or discontinuous (`first-order') transitions.
Coexistence implies an inhomogeneous state spanning separate branches
of the homogeneous flow curves. Note, however, that there is an
ambiguity in connecting separate branches of the homogeneous flow
curves in Fig.~\ref{fig:both}. The top curve permits coexistence of
states with the same stress and different strain rates, while the
lower curve also allows coexistence of states with the same strain
rate and different stresses \cite{schmitt95,olmstedlu97}!

Fig.~\ref{fig:couette} shows that phase separation at a common
stress occurs such that the interface between bands is parallel to the
vorticity-velocity plane (annular bands, in Couette flow), while phase
separation at a common strain rate occurs with the interface
between bands parallel to the velocity--velocity-gradient plane
(stacked discs, in Couette flow).
%%%%%%%%%%%%%%%%%%%%%%%%%%%%%%%%%%%%%%%%%%%%%%%%%%%%%%%%%%%%%%%%%%%
 {\begin{figure}
 \par\columnwidth20.5pc
 \hsize\columnwidth\global\linewidth\columnwidth
 \displaywidth\columnwidth
 \epsfxsize=3.0truein
 \centerline{\epsfbox{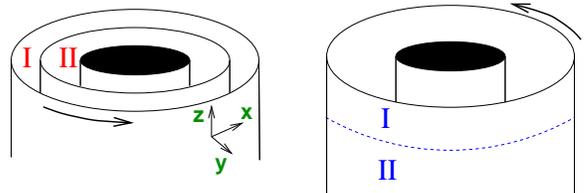}}
 \caption{Geometries for phase separation at common stress (left) or
   strain-rate (right) in a Couette rheometer. For phase separation at
   a common stress (left) phases $I$ and $II$ have different strain
   rates, while at a common strain rate (right) they have different
   stresses. $\hat{\textbf{z}}$ is the vorticity axis,
   $\hat{\textbf{x}}$ is the flow direction, and $\hat{\textbf{y}}$ is
   the flow gradient axis.}
 \label{fig:couette}
 \end{figure}}
%%%%%%%%%%%%%%%%%%%%%%%%%%%%%%%%%%%%%%%%%%%%%%%%%%%%%%%%%%%%%%%%%%%

This highlights a striking contrast between equilibrium and
non-equilibrium systems. In equilibrium the field variables (pressure,
temperature, chemical potential) are uniquely defined and determine
phase coexistence. In sheared fluids, one needs an extra field
variable to determine the extended phase diagram. However, {\sl for a
  system with more than one choice of coexisting geometry, the
  appropriate field variable may not necessarily be identified a
  priori}.  The complete answer of how to determine (theoretically)
the dynamic field variable is not known.  Of course the nature of the
constitutive relation may help, for example the top curve of the
Fig.~\ref{fig:both} does not allow the strain rate as the field
variable. We will come back to discuss some possible answers to this
interesting problem in subsection \ref{which} (see also
\cite{schmitt95} for other suggestions).

Another important difference from equilibrium systems is evident when,
say assuming the system choose to form shear bands at common stress,
we try to determine at which stress a system shear bands. The
constitutive relations shown in Fig.~\ref{fig:both} are calculated for
homogeneous states, and there is no apparent prescription for
determining the selected banding stress, despite the experimental
evidence for a selected stress.  A similar apparent degeneracy occurs
in first order phase transitions in equilibrium statistical mechanics,
but is easily resolved by demanding that the system minimize its total
free energy, or, equivalently, by appealing to the convexity of the
free energy of the equilibrium thermodynamic systems \cite{israel}.
This leads to equality of field variables between two phases and the
common tangent condition ({\sl e.g.\/} the Maxwell equal areas
construction for liquid-gas coexistence \cite{landaustat1}, or the
equal osmotic pressure condition, aided by equal chemical potential,
in rod suspensions \cite{buining93}).

In the shear band problem, an unambiguous resolution of this
degeneracy is to consider the full {\sl inhomogeneous\/} ({\sl i.e.\/}
non-local in space) equations of motion, and determine phase
coexistence by that choice of field variables (appropriately chosen by
hand) for which there exists a {\sl stationary\/} interfacial solution
to the steady-state differential equations of motion
\cite{olmsted90,spenley96,olmstedlu97}.  For zero stress this
technique reduces, as it should, to minimization of the free energy.
The importance of inhomogeneous terms in fluid equations of motion has
been noted by several groups, who pointed out that the standard fluid
equations can have ill-defined mathematical solutions \cite{elkareh89}
if such terms are not included.  Of course, if the phase diagram
depends sensitively on the form or magnitude of the inhomogeneous
terms one need a detail understanding of the underlying physics. The
use of a stable interface to select among possible coexisting states
was first postulated for non-linear dynamical systems, as far as we
know, by Kramer \cite{kramer81a}, and later by Pomeau \cite{pomeau86};
and first applied (independently) to complex fluids in
Ref.~\cite{olmsted92}. The inclusion of gradient terms in constitutive
relations is rapidly gaining acceptance, as recent preprints by Goveas
(phase separation of model blends of long and short polymers)
\cite{goveas99} and Dhont (introduction of model gradient terms to
resolve stress selection) \cite{dhont99} indicate.

In this work we study a model for rigid-rod suspensions in shear flow.
While there are certainly ongoing experiments on these systems
\cite{mather97}, the primary motivation for this extended work is to
explore the manner in which phase separation and coexistence occurs in
complex fluids in flow.  The approximations used in obtaining our
equations are severe (including a decoupling approximation whose
defects are well-known \cite{CLF95}), and we expect qualitative
agreement at best.  However, this is the first complete study of which
we are aware of non-equilibrium phase separation of a complex fluid in
flow for a concrete model, and we hope it illuminates the
phenomenology of flow-induced phase transitions.

%%%%%%%%%%%%%%%%%%%%%%%%%%%%%%%%%%%%%%%%%%%%%%%%%%%%%%%%%%%%%%%%%%%%%%%%%%%%
\section{Methodology}
%%%%%%%%%%%%%%%%%%%%%%%%%%%%%%%%%%%%%%%%%%%%%%%%%%%%%%%%%%%%%%%%%%%%%%%%%%%%
We seek the equations of motion for a solution of rod-like particles.
The most useful dynamic variables describing the long-wavelength
hydrodynamic degrees of freedom are the volume fraction $\phi({\bf
  r})$, the fluid velocity ${\bf v}({\bf r})$, and the nematic order
parameter tensor 
\begin{equation}
Q_{\alpha\beta}({\bf r}) = \langle \nu_{\alpha}\nu_{\beta} - 
\case13\delta_{\alpha\beta}\rangle,
\end{equation}
where $\boldsymbol{\nu}$ is the rod orientations, $\langle\cdot\rangle$ denotes an average around the point ${\bf r}$.   Previous
studies of liquid crystals under shear flow have been either for
thermotropics \cite{olmsted90,olmsted92}, where the issues we
present below associated with composition coupling are not present;
or homogeneous suspensions \cite{see90,bhave93}, 
where phase coexisting was not considered.

Our work below is based on the model extending that of Doi \cite{doi81,doikuzuu83}. 
See, \emph{et al.}  \cite{see90} studied the Doi model in shear flow,
but did not attempt to consider phase coexistence. Bhave, \emph{et al}
\cite{bhave93} analyzed this model in more detail, but did not
consider realistic phase separation behavior. We augment this model
with reasonable estimates for translational entropy loss upon phase
separation and for the free energy cost due to spatial
inhomogeneities.  Zubarev studied shear-induced phase separation in a
variation of the Doi model in flow based on the equality of
non-equilibrium free energies, calculated from the flow-perturbed
orientational distribution function \cite{Zuba96}.  Zubarev only
considered phase separation at a common strain rate, and did not treat
the rheological response (stress) of the system or log-rolling states.

%%%%%%%%%%%%%%%%%%%%%%%%%%%%%%%%%%%%%%%%%%%%%%%%%%%%%%%%%%%%%%%%%%%%%%%%%%%%
\subsection{Equations of Motion}
%%%%%%%%%%%%%%%%%%%%%%%%%%%%%%%%%%%%%%%%%%%%%%%%%%%%%%%%%%%%%%%%%%%%%%%%%%%%
The free energy ({\sl e.g.\/} as in Ref.~\cite{buining93}) is given by 
\begin{eqnarray}
{\cal F\/}(\phi,\boldsymbol{Q})&& =k_{\scriptscriptstyle B}T \int\!d^3\!r
\left\{\phantom{\bigl\{}\!\!\!{\phi\over v_r}\log\phi +
{\left(1\!-\!\phi\right)\over v_s}\log\left(1\!-\!\phi\right) 
\right. \nonumber\\ 
&&
+ {\phi\over v_r} 
\left[ \case12 \left(1\!-\!\case13 u\right)\hbox{\rm Tr}
\,\boldsymbol{Q}^2\!
-\case13 u \hbox{\rm Tr}\,\boldsymbol{Q}^3\!
+\case14 u\left(\hbox{\rm Tr}\,\boldsymbol{Q}^2\right)^2 
\right.
\nonumber\\
&& 
\left.\left. + \case12 K\left(\nabla_{\alpha}Q_{\beta\lambda}\right)^2 \right]
 + \case12 \frac{g}{v_s} \left(\nabla\phi\right)^2 \right\}.
\label{eq:free}
\end{eqnarray}
Here, ``${\rm Tr}$'' denotes the trace, $v_r$ and $v_s$ are rod and
solvent monomer volumes and
\begin{equation}
u\equiv\nu_2 c d L_0^2,
\end{equation}
is Doi's excluded volume parameter \cite{doi81,doikuzuu83}, where $c$
is the concentration (number/volume) of rods of length $L_0$ and
diameter $d$, and $\nu_2$ is a geometrical prefactor
(Ref.~\cite{doi81} estimated $\nu_2=5\pi/16\simeq0.98$). The 
 volume fraction $\phi$ is
\begin{equation}
  \label{eq:phi}
  \phi = c v_r,
\end{equation}
in terms of which
\begin{equation}
u=\phi L \frac{\nu_2}{\alpha},
\end{equation}
where $L=L_0/d$ is the rod aspect ratio and $\alpha$ is an ${\cal
  O\/}(1)$ prefactor defined by
\begin{equation}
  v_r = \alpha d^2L_0.
\end{equation}
For spherocylinders, $\alpha=\pi [1 - 1/(3L)] / 8$ which reduces, in
the limit $L\rightarrow\infty$, to $\alpha=\pi/8\simeq0.39$.  We 
use $u$ and $\phi$ interchangeably below as a composition
variable.

In much of what follows we make two further assumptions to reduce
the number of parameters in our model. We fix  $v_s$ 
by assuming 
\begin{equation}
  v_r = L v_s,
\label{eq:approx1}
\end{equation}
which corresponds to a particular volume of the solvent molecules
relative to that of the rod-like molecules.  Further, we assume that
the geometric factor $\nu_2/\alpha$ has the value unity, so that
\begin{equation}
  u = \phi L,
\label{eq:approx2}
\end{equation}
which corresponds to a particular shape of the rigid-rod molecules.
These two assumptions specify the detailed shape and volume ratio of
the system we study below. For slightly different systems with $v_r
\ne L v_s$ or $\nu_2/\alpha \ne 1$, our work should still provide an
accurate qualitative picture.

The first two terms of Eq.~(\ref{eq:free}) comprise the entropy of
mixing, and the first three terms in square brackets are from Doi's
expansion of the free energy (derived per solute molecule) in powers
of the nematic order parameter $\boldsymbol{Q}$. These terms were
derived from the Smoluchowski equation for the distribution function
of rod orientations \cite{doi81,doikuzuu83}. We keep the expansion to
fourth order to describe a first order transition and give the correct
qualitative trends.

Assuming Eqs.~(\ref{eq:approx1}-\ref{eq:approx2}),
we calculate the following biphasic coexistence regions,
\begin{eqnarray}
  \{u_{I}=2.6925, u_{N}=2.7080\} &&\qquad (L=5.0)\label{eq:INa} \\
  \{u_{I}=2.6930, u_{N}=2.7074\} &&\qquad (L=4.7),\label{eq:INb}
\end{eqnarray}
where $u_I$ and $u_N$ are the excluded volume parameters
(compositions) for the coexisting isotropic and nematic phases,
respectively. Note the very weak dependence of the biphasic regime (in
the scaled variable $u=L\phi$) on $L$.

The last two terms in Eq.~(\ref{eq:free}) penalize spatial
inhomogeneities.  By adding the single term proportional to $K$ we
have assumed a particular relation for the Frank constants,
($K_1\!=\!K_2\!=\!K, K_3\!=\!0)$ \cite{lubensky70,pgdg}.  Although
Odijk has calculated these constants for model liquid crystals (in the
nematic regime) \cite{odijk86}, we will see below that this choice is
probably unimportant for this model. More generally, we expect the
Frank constants to vary as functions of $\boldsymbol{Q(r)}$ in
physical systems, a situation which we have not addressed here. The
final term penalizes composition gradients \cite{gunton}.  We are not
aware of any calculations of $g$ for solutions of rod-like particles.
In Eq.~(\ref{eq:free}), we assume an athermal solution with no
explicit interaction energy.

The nematic order parameter obeys the following equation of motion
\cite{doi81,doikuzuu83}:
\begin{equation}
\left(\partial_t + {\rm\bf v}\!\cdot\!\boldsymbol{\nabla}\right) \boldsymbol{Q} =
\boldsymbol{F}(\boldsymbol{\kappa},\boldsymbol{Q}) + \boldsymbol{G}
(\phi, \boldsymbol{Q})  
\label{eq:2} 
\end{equation}
where $\kappa_{\alpha\beta} = \nabla_{\beta} v_{\alpha}$.  In
Eq.~(\ref{eq:2}) the (reactive) ordering term $\boldsymbol{F}$ is given by
\begin{equation}
  \boldsymbol{F}(\boldsymbol{\kappa},\boldsymbol{Q})\!=\!\case23\boldsymbol{\kappa}^{s}\!+ 
  \!\boldsymbol{\kappa}\!\cdot\!\boldsymbol{Q}\!+\!
  \boldsymbol{Q}\!\cdot\!\boldsymbol{\kappa}^{\scriptscriptstyle T}
  \!-\!2(\boldsymbol{Q}\!+\!\case13\boldsymbol{I})\,\hbox{\rm
    Tr}(\boldsymbol{Q}\!\cdot\!\boldsymbol{\kappa}),\label{eq:convect}
\end{equation}
where $\boldsymbol{\kappa}^s$ is the symmetric part of
$\boldsymbol{\kappa}$ and $\boldsymbol{I}$ is the identity tensor.
For simplicity, we have chosen the form appropriate for an infinite
aspect ratio (the prefactors differ by ${\cal O\/}(1)$ constants for
finite aspect ratios \cite{doikuzuu83}).  The coupling
$\boldsymbol{F}$ to the flow both induces order, and dictates a
preferred orientation. The dissipative portion $\boldsymbol{G}$ is
\begin{equation} 
\boldsymbol{G}(\phi,\boldsymbol{Q}) = 6 \frac{ \bar{D}_{\mit
  r}}{k_{\scriptscriptstyle B}T}\frac{v_r}{\phi} \boldsymbol{H},
\label{eq:G}
\end{equation} 
where
\begin{equation}
  \label{eq:Dr}
  \bar{D}_r=\frac{\nu_1 D_{r0}}{(1-\case32\,\hbox{\rm Tr}\,\boldsymbol{Q}^2)^2
    (c L_0^3)^2},
\end{equation}
is the collective rotational diffusion coefficient and

\begin{equation}
\boldsymbol{H} = -\left[\frac{\delta{\cal F\/}}{\delta \boldsymbol{Q}}
-\case13\,\boldsymbol{I}\,\textrm{Tr}\frac{\delta{\cal F\/}}{\delta
  \boldsymbol{Q}}
\right]
\end{equation}
is the molecular field. $D_{\mit r0}$ is the single-rod rotational
diffusion coefficient and $\nu_1$ is an ${\cal O\/}(1)$ geometrical
prefactor, which will be fixed below Eq.~3.30.
The rotational diffusion coefficient is
\begin{equation}
  \label{eq:Dr0}
  D_{r0} = \frac{k_{\scriptscriptstyle B}T\ln L}{3\pi\eta L_0^3},
\end{equation}
where $\eta$ is the solvent viscosity. The $\boldsymbol{Q}$-dependence
in the denominator of Eq.~(\ref{eq:Dr}) enhances reorientation for
well-ordered systems \cite{doi81}. Our choice for $\bar{D}_r$ is
crude, since it applies to rods in concentrated solution and we use it
in the concentrated and semi-dilute regimes.  As with many of our
approximations, this gives us a tractable model system with which to
study the phenomenology of phase separation.

Doi and co-workers derived Eq.~(\ref{eq:2}) for homogeneous systems.
We extend this to inhomogeneous systems by including the gradient
terms implicit in the functional derivative which defines
$\boldsymbol{H}$.  Our choice of $\boldsymbol{F}$ is the so-called
quadratic closure approximation to the Smoluchowski equation
\cite{doikuzuu83}.  This approximation ensures that the magnitude of
the order parameter remain in the physical range in the limit of
strong ordering, but is known to incorrectly predict phenomena such as
director tumbling and wagging. Many workers have investigated the
subtleties of various closure approximations and the degree to which
they reproduce realistic flow behavior \cite{CLF95}. Since our primary
goal is to explore the method for calculating phase behavior and
outline some of the possibilities for coexistence under flow, we
confine ourselves to this well-studied model.

The fluid velocity obeys \cite{doi81,doikuzuu83,lulu}:
\begin{equation}
\rho\left(\partial_t + {\rm\bf v}\cdot\boldsymbol{\nabla}\right) {\rm\bf v} =
\boldsymbol{\nabla}\!\cdot\!\left[2\eta\boldsymbol{\kappa}^s +
\boldsymbol{\sigma} (\phi, \boldsymbol{\kappa},\boldsymbol{Q}) \right]
+ \frac{\delta\cal F}{\delta\phi}\boldsymbol{\nabla}\!\phi
-\boldsymbol{\nabla} p,
\label{eq:1}
\end{equation}
where $\eta$ is the solvent viscosity, $\rho$ the fluid mass density,
and the pressure $p$ enforces incompressibility,
$\boldsymbol{\nabla}\!\cdot\!{\rm\bf v}=0$.  For the low Reynold's
number situations considered here, and for steady shear flow, we will
equate the left-hand side of the equation above to zero.

The constitutive relation for the stess tensor $\boldsymbol{\sigma} (\phi,
\boldsymbol{\kappa},\boldsymbol{Q})$ was derived by Doi and co-workers, and
includes dissipative and elastic parts. Since the elastic stress
dominates \cite{doiedwards}, we keep only this part:
\begin{eqnarray}
\boldsymbol{\sigma}&\simeq& \boldsymbol{\sigma}_{\mit elastic} \nonumber \\
&=& - 3\boldsymbol{H} + 
\boldsymbol{H}\!\cdot\!\boldsymbol{Q} - \boldsymbol{Q}\!\cdot\!\boldsymbol{H} 
- \boldsymbol{\nabla}Q_{\alpha\beta}\cdot
{\delta{\cal F\/}\over\delta\boldsymbol{\nabla}Q_{\alpha\beta}}.
\label{eq:stress}
\end{eqnarray}
The first term of Eq.~(\ref{eq:stress}) was given by Doi \cite{doi81},
while the last three terms were derived later \cite{olmsted90} and are
equivalent to the elastic stress due to Frank elasticity \cite{pgdg},
generalized to a description in terms of the nematic order parameter
$\boldsymbol{Q}$ rather than the nematic director. Note that the last
three terms vanish for a homogeneous system.

Finally, the composition equation of motion is of the Cahn-Hilliard
form \cite{gunton};
\begin{align}
\left(\partial_t + {\rm\bf v}\cdot\boldsymbol{\nabla}\right) \phi &=
- \boldsymbol{\nabla}\cdot\boldsymbol{J} \nonumber\\
&=\boldsymbol{\nabla}\!\cdot\!\boldsymbol{M}\!\cdot\!\boldsymbol{\nabla} \mu,
\label{eq:3}
\end{align}
where $\boldsymbol{M}$ is the mobility tensor and the chemical potential is
given by
\begin{equation}
\mu = \frac{\delta {\cal F\/}}{\delta\phi}.
\label{eq:mu0}
\end{equation}
The diffusive current is $\boldsymbol{J} =
-\boldsymbol{M}\!\cdot\!\boldsymbol{\nabla}\mu$.  The complete
dynamics is thus described by Eqs.~(\ref{eq:2},\ref{eq:1})
and~(\ref{eq:3}).

The dynamical equations of motion for other complex fluids have the
same theoretical structure: equations of motion for the conserved
quantities and the broken-symmetry or flow-induced structural order
parameter (analogous to $\boldsymbol{Q}$), and a constitutive relation
for the stress as a function of composition and order parameter
\cite{lulu}. For a given system and set of equations of motion, the
analysis below is generic.  For some local models, internal dynamics
(Eq.~\ref{eq:2}) can be eliminated to give the stress as a history
integral over the strain rate. In polymer melts \cite{doiedwards}, and
in wormlike micelles \cite{cates90} far from a nematic regime, this
leads to non-monotonic stress--strain-rate curves. However, augmenting
these integral theories with non-local terms to calculate interface
profiles is non-trivial.

%%%%%%%%%%%%%%%%%%%%%%%%%%%%%%%%%%%%%%%%%%%%%%%%%%%%%%%%%%%%%%%%%%%%%%%%%%%%
\subsection{Steady-state conditions}
%%%%%%%%%%%%%%%%%%%%%%%%%%%%%%%%%%%%%%%%%%%%%%%%%%%%%%%%%%%%%%%%%%%%%%%%%%%%
In this work we study planar shear flow, specified by
\begin{equation}
\frac{\partial v_x({\rm\bf r})}{\partial y}  = \dot{\gamma}(\textbf{r}).
\end{equation}
For homogeneous flows ${\rm\bf v}({\rm\bf r}) = \dot{\gamma} y
{\rm\bf\hat{x}}.$  
The phase diagram is given by the domains of stable
steady-state solutions to the equations of motion for applied shear
stress or strain-rate, in the phase space spanned by
\begin{align}
({\phi},{\sigma}_{xy}),&&&\text{(common stress)}\\
({\phi},{\dot{\gamma}}),&&&\text{(common strain rate).}
\end{align}
For phase separation at common stress the stress is uniform and the
strain rate partitions between the two phases; while for phase
separation at common strain rate the strain rate is uniform and the
shear stress partitions between the two phases.

The strain rate tensor is given by
\begin{equation}
\boldsymbol{\kappa} = \dot{\gamma}\left(
\begin{array}{ccc}
0 & 1 & 0 \\
0 & 0 & 0 \\
0 & 0 & 0 
\end{array} \right).
\end{equation}

Upon rescaling,
\begin{eqnarray}
\widehat{\dot{\gamma}} &=& \frac{\dot{\gamma} L^2}{6D_{\mit r0}\nu_1\nu_2^2} \\
\widehat{\boldsymbol{\sigma}} &=&\frac{\sigma\nu_2 L^3}{3
  k_{\scriptscriptstyle B} T},
\end{eqnarray}
the steady-state condition for the order parameter (Eq.~\ref{eq:2}) is
\begin{equation}
0 = \frac{1}{u^2L^2(1-\case32\hbox{\rm Tr}\boldsymbol{Q}^2)^2} 
\widehat{\boldsymbol{H}}
+ \widehat{\dot{\gamma}}\,\widehat{\boldsymbol{F}},
\label{eq:Q}
\end{equation}
where $\boldsymbol{F} = \dot{\gamma} \widehat{\boldsymbol{F}}$ and 
\begin{equation}
  \label{eq:H}
-\widehat{\boldsymbol{H}} = \left(1\!-\!\frac{\scriptstyle u}{\scriptstyle
    3}\right)
\boldsymbol{Q} - u\left(\boldsymbol{Q}^2\!-\!
  \frac{\boldsymbol{I}}{3}\textrm{Tr}\boldsymbol{Q}\right) +u
\boldsymbol{Q}\textrm{Tr}\boldsymbol{Q}^2 - K\nabla^2\boldsymbol{Q}.
\end{equation}

In steady state planar shear flow the velocity gradients are normal to
the flow direction, so the convective derivative vanishes and
Eq.~(\ref{eq:Q}) specifies the order parameter in a homogeneous flow.
Under these conditions integration of the momentum equation
Eq.~(\ref{eq:1}) gives a constant stress,
\begin{equation}
  \boldsymbol{\sigma}_0=\boldsymbol{\sigma} - p\,\boldsymbol{I} + 2 \eta\boldsymbol{\kappa}^s,
\end{equation}
where $\boldsymbol{\sigma}_0$ is the boundary stress.  The rescaled
shear stress is
\begin{equation}
\widehat{\sigma}_{xy}^0 = A \widehat{\dot{\gamma}} - u L \left[
\widehat{\boldsymbol{H}}+ K \left(\nabla^2\boldsymbol{Q}\!\cdot\!\boldsymbol{Q} -
\boldsymbol{Q}\!\cdot\!\nabla^2\boldsymbol{Q}\right)\right]_{xy},
\label{eq:sigma}
\end{equation}
where $A=2\nu_1\nu_2^3(\ln L)/(3\pi)$ is a constant of order unity: we
take $A=1$ for the remainder of this work, which corresponds to a
particular choice for $\nu_1$.  As with the assumptions of molecular
geometry embodied in $\nu_2$ and $\alpha$
(Eqs.~\ref{eq:approx1}-\ref{eq:approx2}), different values for $A$
should not qualitatively change the nature of our results.

Integrating the steady-state composition equation (\ref{eq:3}) and
using the boundary condition that material cannot enter or leave the
system, we find
\begin{eqnarray}
  \mu_0 &=& \mu({\rm\bf r}) \label{eq:mu} \\
  \frac{\mu(\textbf{r})}{k_{\scriptscriptstyle B}T} &=&
  F_{\mit Doi} + \frac{\partial}{\partial\phi}\left[\phi\ln\phi + L
    \left(1-\phi\right)\ln\left(1-\phi\right)\right]  \label{eq:mucalc} 
\nonumber \\
  && + \phi L \frac{\partial}{\partial u}
  F_{\mit Doi} + \case12K \left(\nabla\boldsymbol{Q}\right)^2 - g L
  \nabla^2\phi
  \label{eq:mu1}
\end{eqnarray}
where Eqs.~(\ref{eq:approx1}-\ref{eq:approx2}) have been used to
specify the molecular geometry,  $\mu_0$ is a constant of
integration, and

\begin{equation}
  \label{eq:Fdoi}
F_{Doi} = \case12 \left(1\!-\!\case13 u\right)\hbox{\rm Tr}
\,\boldsymbol{Q}^2\!
-\case13 u \hbox{\rm Tr}\,\boldsymbol{Q}^3\!
+\case14 u\left(\hbox{\rm Tr}\,\boldsymbol{Q}^2\right)^2.
\end{equation}
Note that the mobility tensor $\boldsymbol{M}$ plays no role in the
steady-state conditions, or in the resulting phase diagram.  

Eqs.~(\ref{eq:Q},\ref{eq:sigma}) and~(\ref{eq:mu}) completely specify
the system in planar shear.  Solving these equations will occupy the
remainder of this work. Note that variables $\widehat{\dot{\gamma}},
\widehat{\boldsymbol{\sigma}},$ $\mu/k_{\scriptscriptstyle B} T$ are
all dimensionless quantities.

%%%%%%%%%%%%%%%%%%%%%%%%%%%%%%%%%%%%%%%%%%%%%%%%%%%%%%%%%%%%%%%%%%%%%%%%%%%%
\section{Calculation of Phase Diagrams}
%%%%%%%%%%%%%%%%%%%%%%%%%%%%%%%%%%%%%%%%%%%%%%%%%%%%%%%%%%%%%%%%%%%%%%%%%%%%
\subsection{Interface Calculation}
%%%%%%%%%%%%%%%%%%%%%%%%%%%%%%%%%%%%%%%%%%%%%%%%%%%%%%%%%%%%%%%%%%%%%%%%%%%%

The phase diagram is specified by solving
Eqs.~(\ref{eq:Q},\ref{eq:sigma},\ref{eq:mu}) for given $\mu_0$ and
boundary stress $\sigma_{xy}^0$.  Non-equilibrium `phases' are defined
as the stable steady-state space-uniform solutions to these equations.
These {\sl inhomogeneous\/} equations comprise a set of ordinary
differential equations, through the gradients that appear in the
stress and in the functional derivatives that define $\mu$ and
$\boldsymbol{H}$. The only parameters of the theory are the rod aspect
ratio $L$ and the ratio of elastic constants,
\begin{equation}
\lambda=\frac{gL}{K}
\end{equation} 
($K$ may be absorbed into the length scale of the system).

%%%%%%%%%%%%%%%%%%%%%%%%%%%%%%%%%%%%%%%%%%%%%%%%%%%%%%%%%%%%%%%%%%%
 {\begin{figure}
 \par\columnwidth20.5pc
 \hsize\columnwidth\global\linewidth\columnwidth
 \displaywidth\columnwidth
 \epsfxsize=3.5truein
 \centerline{\epsfbox[70 230 670 650]{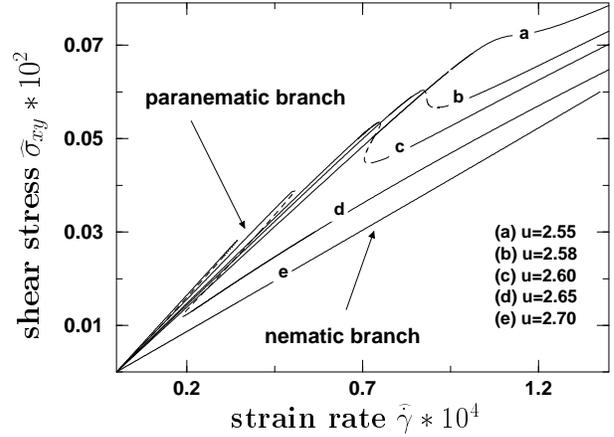}}
 \caption{Homogeneous stress $\hat{\sigma}_{xy}$ vs. strain rate
   $\widehat{\dot{\gamma}}$ behavior for various excluded volumes,
   $L=5.0$ and $\lambda=1.0$.  Dotted lines mark unstable branches.
   Similarly, curves along which $\partial\mu/\partial\phi<0$ are
   linearly unstable.}
 \label{fig:constit}
 \end{figure}}
%%%%%%%%%%%%%%%%%%%%%%%%%%%%%%%%%%%%%%%%%%%%%%%%%%%%%%%%%%%%%%%%%%%
We first fix $\phi$ ($u$) and solve the homogeneous algebraic versions
of Eqs.~(\ref{eq:Q}) and~(\ref{eq:sigma}) for $\boldsymbol{Q}$ and
$\dot{\gamma}$ as a function of $\sigma_{xy}^0$\footnote{In the few
  cases where the phase diagram in the $\sigma_{xy}\!-\!\mu$ plane has
  a transition line parallel to the $\mu$ axis, one must first fix
  $\mu_0$, and then determine $\sigma_0$.}.  This is done for all
$\phi$. Because ${\cal F\/}(\phi,\boldsymbol{Q})$ describes an I-N
transition, at a given stress, multiple roots exist, with distinct
strain rates and $\boldsymbol{Q}$.  Fig.~\ref{fig:constit} shows the
stress strain-rate relations for homogeneous solutions to
Eqs.~(\ref{eq:Q}) and~(\ref{eq:sigma}) for $L=5.0$ and $\lambda=1.0$.

%%%%%%%%%%%%%%%%%%%%%%%%%%%%%%%%%%%%%%%%%%%%%%%%%%%%%%%%%%%%%%%%%%%
 {\begin{figure}
 \par\columnwidth20.5pc
 \hsize\columnwidth\global\linewidth\columnwidth
 \displaywidth\columnwidth
 \epsfxsize=3.5truein
 \centerline{\epsfbox[0 0 712 370]{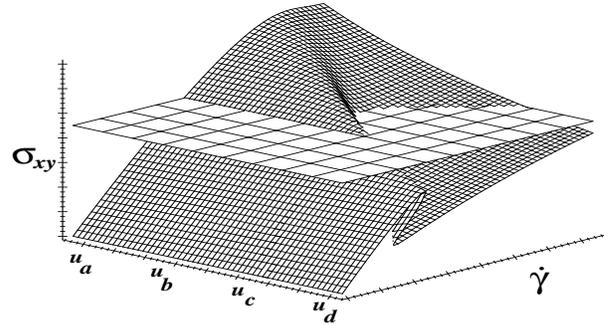}}
 \caption{Stress-strain-composition surface for the curves in
   Fig~\protect{\ref{fig:constit}}. The plane is at
   $\widehat{\sigma}_{xy}^0=0.05$}
 \label{fig:3D}
 \end{figure}}
%%%%%%%%%%%%%%%%%%%%%%%%%%%%%%%%%%%%%%%%%%%%%%%%%%%%%%%%%%%%%%%%%%%
The isotropic branch has a larger viscosity than the nematic branch,
and has an increasing effective viscosity for increasing
concentration, reflecting the contribution $u\widehat{\boldsymbol{H}}$
in Eq.~(\ref{eq:sigma}). Conversely, the nematic branch has a lower
stress at higher concentrations due to the increased nematic order
which permits less-hindered motion.

For a dilute isotropic system (curve \textsf{a}), shear flow
continuosly induces nematic order. A more aligned system has a lower
effective viscosity, so the stress $\sigma(\dot{\gamma})$ increases
slower than linearly (shear-thins) as the magnitude of the order
parameter $\boldsymbol{Q}$ increases. Eventually the system attains,
smoothly, a high strain rate state with a much lower viscosity than in
the limit of zero stress.  For more concentrated systems (curves
\textsf{b}, \textsf{c} and \textsf{d}) shear flow induces a transition
to a nematic phase with lower viscosity, and $\sigma(\dot{\gamma})$ is
non-monotonic.  There is a region of stresses for which two stable
strain rates exist, on either the nematic or isotropic branches of the
constitutive curve.  For compositions inside the biphasic regime
(curve \textsf{e}) both nematic and isotropic branches exist in the
limit of zero stress, with the isotropic branch losing stability at
high enough stress. Finally (not shown) for highly concentrated
systems only nematic branches exist.

As mentioned in Sec.~IIB, we calculate the phase diagram by explicitly
constructing the coexisting interfacial solution \cite{olmsted92}.  In
common stress coexistence, for example, the coexisting states have
{\sl different\/} strain rates and, generally, different compositions.
Hence, they connect the high and low strain rate branches of two
different curves in Fig.~\ref{fig:constit}. It is easiest to visualize
this by considering the intersection of a plane at a given stress
$\sigma_{xy}^0$ with the surface $\sigma_{xy}(\dot{\gamma},u)$, as in
Fig.~\ref{fig:3D}.
%%%%%%%%%%%%%%%%%%%%%%%%%%%%%%%%%%%%%%%%%%%%%%%%%%%%%%%%%%%%%%%%%%%
 {\begin{figure}
 \par\columnwidth20.5pc
 \hsize\columnwidth\global\linewidth\columnwidth
 \displaywidth\columnwidth
 \epsfxsize=3.25truein
 \centerline{\epsfbox[150 220 495 660]{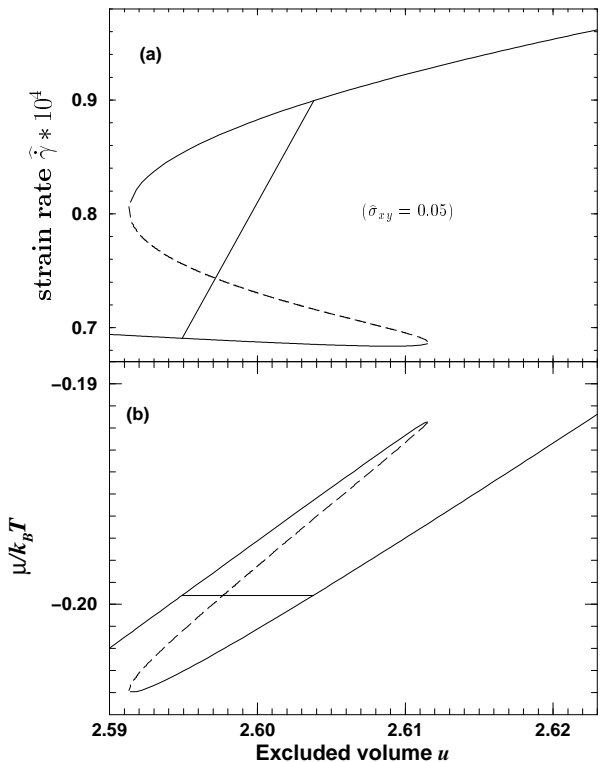}}
 \caption{(a) Reduced strain rate $\widehat{\dot{\gamma}}(u)$ and
   (b) chemical potential $\mu(u)$ for the stress contour in
   Fig.~\protect{\ref{fig:3D}} ($\widehat{\sigma}_{xy}=0.05$). The tie
   line is calculated using the interface construction.}
 \label{fig:strainmu}
 \end{figure}}
%%%%%%%%%%%%%%%%%%%%%%%%%%%%%%%%%%%%%%%%%%%%%%%%%%%%%%%%%%%%%%%%%%%

At a given stress, the strain rate varies with composition as shown in
Fig.~\ref{fig:strainmu}a. At coexistence, the chemical potential
$\mu({\rm\bf r})$ must be constant through the interface, as dictated
by Eq.~(\ref{eq:mu}). The functional form of the non-equilibrium
chemical potential is known from Eq.~(\ref{eq:mu1}), and depends on
the strain rate through the dependence of the nematic order parameter
on the strain rate in steady state.  We plot $\mu(u)$ in
Fig.~\ref{fig:strainmu}b. There is a continuum range of $\mu$, which
allow possible coexisting pairs of states.  (Recall that $u$ is
proportional to the rod volume fraction $\phi$).

We now impose the interface solvability condition as follows.  For a
given stress $\sigma_{xy}^0$, we determine a specific coexistence
chemical potential $\mu_0$, which allows a stable interfacial solution
to Eqs.~(\ref{eq:Q},\ref{eq:sigma},\ref{eq:mu}).  In practice, we
eliminate $\dot\gamma({\rm\bf r})$ from Eq.~(\ref{eq:Q}) using
Eq.~(\ref{eq:sigma}), and solve Eqs.~(\ref{eq:Q}) and~(\ref{eq:mu})
for the interfacial profile, with boundary conditions (fixed
$\boldsymbol{Q}$ and $u$) chosen by two points on the low and high
strain rate branches of Fig.~\ref{fig:strainmu}b with the same
$\mu_0$. We adjust $\mu_0$ until a stationary interfacial profile is
found. This solvability criterion give sharp selection on $\mu$, and
in this way determine a tie line on the $\dot{\gamma}\!-\!u$ plane,
Fig.~\ref{fig:strainmu}a. By varying the stress we compute the entire
phase diagram in the $\sigma_{xy}-u$ and $\dot{\gamma}-u$ planes.

For phase separation at a common strain rate the construction is
analogous. One slices a vertical plane through Fig.~\ref{fig:3D} at a
given strain rate, constructs the curve $\mu(u)$ along the
intersection with the surface, and searches for a stationary
interfacial solution.

The interface calculations are carried out by discretizing the system
on a one-dimension mesh and, from smooth initial conditions, evolving
Eqs.~(\ref{eq:Q},\ref{eq:sigma},\ref{eq:mu}) forward using fictitious
dynamics calculated with an implicit Crank-Nicholson scheme.  Spatial
variations are only allowed in the direction in which phase separation
occurs, so we replace
\begin{equation}
  \label{eq:5}
  \nabla \rightarrow \begin{cases}
    \dfrac{\partial}{\partial y} & \hbox{common stress} \\[10truept]
    \dfrac{\partial}{\partial z} & \hbox{common strain rate,} 
  \end{cases}
\end{equation}
where $z$ is in the vorticity direction.  We fix the values of $u$ and
$\boldsymbol{Q}$ at either side of the interface to lie on the high
and low strain rate branches of Fig.~\ref{fig:strainmu}, begin with
smooth initial conditions, and let the system ``evolve'' towards
steady state. An interface develops between the two phases, and moves
to one boundary or the other. For a given stress, coexistence is
determined by that chemical potential $\mu$ for which a stationary interface
lies in the interior of the system (in the limit of large system size)
\cite{olmsted92}.  An analogous construction may be made by
maintaining a fixed mean strain rate on the unstable part of a
homogeneous curve, and then starting up the system and allowing it to
select a stress and chemical potential. In either case the selected
stress is that stress for which a stationary interfacial solution
between the high and low strain rate branches \emph{exists}.  Such an
interfacial solution is known in dynamical systems theory as a
heteroclinic orbit \cite{kramer81a,pomeau86,olmstedlu97}, and further
work will investigate this in more detail for simpler model systems
\cite{jsplanar,history,orpdo}.

We restrict the nematic order parameter to
\begin{equation}
\boldsymbol{Q}=\left(
\begin{array}{ccc}
q_1 & q_3 & 0 \\
q_3 & q_2 & 0 \\
0 & 0 & - (q_1 + q_2)
\end{array}
\right),
\end{equation}
since all steady state solutions with non-zero elements $Q_{xz}$ or
$Q_{yz}$ are unstable due to the symmetry of shear flow
\cite{bhave93}. In a similar calculation for thermotropic nematics in
shear flow we have found that this restriction on $\boldsymbol{Q}$
reproduces the same selected stress as that obtained when keeping the
full tensor \cite{olmsted92}.

For planar shear flow and a wide class of equations of motion we have
shown that, if a coexisting solution exists, it occurs at discrete
points in the parameter set\cite{jsplanar}. For example, for a given
stress $\sigma_{xy}^0$, coexistence can occur only at discrete values
for $\mu_0$, that is, along lines in the field variable space spanned
by $\sigma_{xy}\!-\!\mu$. This is analogous to equilibrium systems
where, for example, phase transitions in a simple fluid occur along
lines, rather than within regions, in the pressure-temperature plane.

Note that a one-dimensional calculation does not determine the
stability of the interfacial solution with respect to transverse
undulations (capillary waves), which could be important in,
particularly, the common stress geometry \cite{renardy}. 

%%%%%%%%%%%%%%%%%%%%%%%%%%%%%%%%%%%%%%%%%%%%%%%%%%%%%%%%%%%%%%%%%%%%%%%%%%%%
\subsection{Homogeneous Solutions}
%%%%%%%%%%%%%%%%%%%%%%%%%%%%%%%%%%%%%%%%%%%%%%%%%%%%%%%%%%%%%%%%%%%%%%%%%%%%
The modified Doi model in the quadratic closure approximation has
three stable solutions in homogeneous planar shear flow. We refer the
reader to Bhave, {\sl et al.\/} for further details \cite{bhave93}.

\begin{itemize}
\item[{\sf I}] \emph{Paranematic:} The paranematic state {\sf I}
  induced from a disordered equilibrium phase. The order parameter
  $\boldsymbol{Q}$ is small and fairly biaxial, with major axis lying
  in the shear plane at an angle of almost $\pi/4$ relative to the
  flow direction.
\item[\textsf{N}] \emph{Flow-Aligning Nematic:} The flow-aligning
  nematic state is much more strongly-aligned, has slight biaxiality
  induced by the flow, and has the major axis of alignment in the flow
  plane at an angle of a few degrees relative to the flow direction.
  The \textsf{I} and \textsf{N} states have the same symmetry.
\item[\textsf{L}] \emph{Log-Rolling Nematic:} The log-rolling phase is
  also a well-aligned and almost uniaxial phase, but with major axis of
  alignment in the vorticity $({\rm\bf\hat{z}})$ direction, so the
  rods spin about their major axes.
\end{itemize}

%%%%%%%%%%%%%%%%%%%%%%%%%%%%%%%%%%%%%%%%%%%%%%%%%%%%%%%%%%%%%%%%%
 {\begin{figure}
 \par\columnwidth20.5pc
 \hsize\columnwidth\global\linewidth\columnwidth
 \displaywidth\columnwidth
 \epsfxsize=3.5truein
 \centerline{\epsfbox[40 210 590 640]{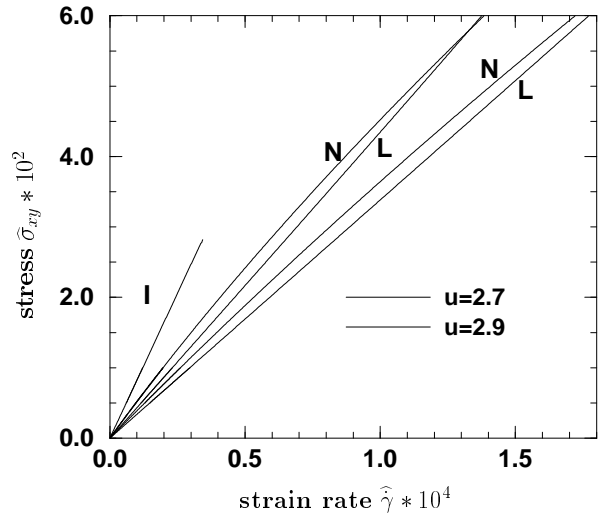}}
 \caption{Constitutive relations for \textsf{I}, \textsf{N}, and
   \textsf{L} states, for $L=5.0$ and two values for the excluded
   volume parameter.}
 \label{fig:log}
 \end{figure}}
%%%%%%%%%%%%%%%%%%%%%%%%%%%%%%%%%%%%%%%%%%%%%%%%%%%%%%%%%%%%%%%%%%%

The {\sf I} phase is stable at lower volume fractions and merges with
the {\sf N} phase at high strain rates. The {\sf L} phase is stable
only at higher volume fractions, and is destabilized at high enough
strain rates. For low strain rates the stress of the {\sf L} state is
lower than that of the {\sf N} state, which is lower than that of the
{\sf I} state (see Fig.~\ref{fig:log}).

Fig.~\ref{fig:stab} shows the regions of stability of the various
states.  The loop in Fig.~\ref{fig:stab}a occurs for compositions such
that the constitutive curve $\sigma(\dot{\gamma})$ has the shape of
curve \textsf{b} in Fig.~\ref{fig:constit}. Similar phase-plane plots
were calculated by Bhave \emph{et al.}  \cite{bhave93} and See
\emph{et al.} \cite{see90}. They did not consider the mixing entropy
needed to generate a realistic nematic transition, however, and always
generated solutions for a given strain rate instead of a given stress.
(this explains the absence of a loop in their phase-plane plot
$\dot{\gamma}\!-\!\phi$). Their plots (compare Fig.~5 of
Ref.~\cite{bhave93}) correspond to truncating the loop in
Fig.~\ref{fig:stab}a.  Fig.~\ref{fig:stab}b has a similar, barely
discernable, loop near the critical point, within which there are no
stable states. This instability is due to the instability of the
composition equation, Eq.~(~\ref{eq:3}). In this region
\begin{equation}
  \label{eq:6}
  \left.\frac{\partial \mu}{\partial \phi}\right|_{\sigma_{\mit xy}} < 0,
\end{equation}
which is equivalent to a negative diffusion coefficient, and is
analogous to the conventional definition of the spinodal line for
ordinary equilibrium demixing.
%%%%%%%%%%%%%%%%%%%%%%%%%%%%%%%%%%%%%%%%%%%%%%%%%%%%%%%%%%%%%%%%%%%
 {\begin{figure}
 \par\columnwidth20.5pc
 \hsize\columnwidth\global\linewidth\columnwidth
 \displaywidth\columnwidth
 \epsfxsize=3.5truein
 \centerline{\epsfbox[150 230 500 640]{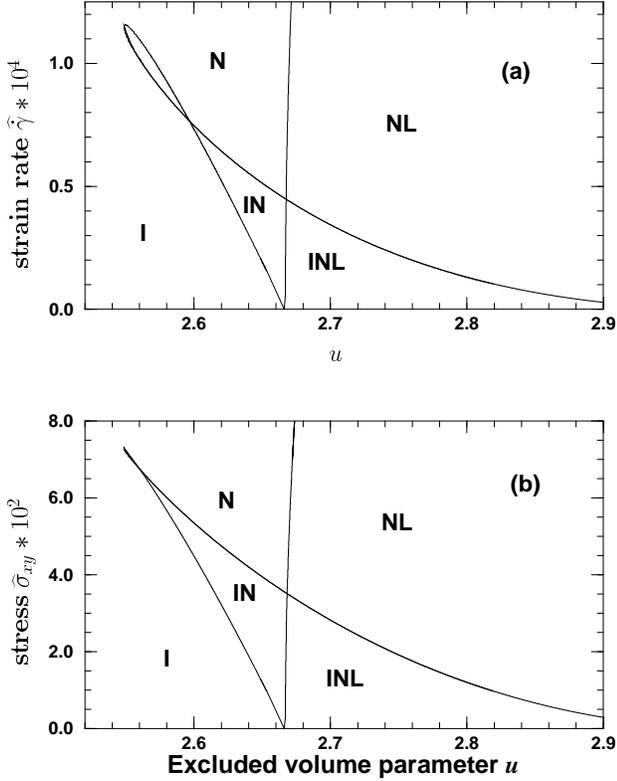}}
 \caption{Regions of stability of paranematic ({\sf I}), 
   nematic ({\sf N}) and log-rolling ({\sf L}) states in the
   strain-rate--composition (a) and stress-composition (b) planes for
   $L=5.0$. Note that the loop in (a) contains \emph{no} stable states.
   Stability limits are calculated with respect to both order parameter
   and composition fluctuations, for a given controlled stress. The
   thin loop in (b) encloses a region with no stable states, due to the
   instability of the composition equation.}
 \label{fig:stab}
 \end{figure}}
%%%%%%%%%%%%%%%%%%%%%%%%%%%%%%%%%%%%%%%%%%%%%%%%%%%%%%%%%%%%%%%%%%%

%%%%%%%%%%%%%%%%%%%%%%%%%%%%%%%%%%%%%%%%%%%%%%%%%%%%%%%%%%%%%%%%%%%%%%%%%%%%
\section{Common Stress Coexistence}
%%%%%%%%%%%%%%%%%%%%%%%%%%%%%%%%%%%%%%%%%%%%%%%%%%%%%%%%%%%%%%%%%%%%%%%%%%%%
For common stress coexistence the interface lies in the
velocity-vorticity plane, and inhomogeneities are in the
${\rm\bf\hat{y}}$ direction (see Fig.~\ref{fig:couette}). The stress
balance condition at the interface is
$\boldsymbol{\sigma}\!\cdot\!{\rm\bf\hat{y}}$ uniform.  $\sigma_{yy}$
is taken care of by the pressure and $\sigma_{zy}$ vanishes by
symmetry (no flow in the ${\rm\bf\hat{z}}$ direction), leaving
continuity of the shear stress $\sigma_{xy}$ through the interface.
The two coexisting phases $I$ and $II$ have  strain rates and
compositions partitioned according to
\begin{eqnarray}
\bar{\phi} &=& \zeta \phi_{\scriptscriptstyle I} + (1-\zeta)
\phi_{\scriptscriptstyle II} \label{eq:lev1}\\
\bar{\dot{\gamma}} &=& \zeta \dot{\gamma}_{\scriptscriptstyle I} + (1-\zeta)
\dot{\gamma}_{\scriptscriptstyle II},\label{eq:lev2}
\end{eqnarray}
where $\bar{\phi}$ and $\bar{\dot{\gamma}}$ are the mean composition
and strain rate and $\zeta$ is the fraction of material in phase $I$.

%%%%%%%%%%%%%%%%%%%%%%%%%%%%%%%%%%%%%%%%%%%%%%%%%%%%%%%%%%%%%%%%%%%%%%%%%%%%
\subsection{Paranematic--flow aligning coexistence ({\sf I-N})}
%%%%%%%%%%%%%%%%%%%%%%%%%%%%%%%%%%%%%%%%%%%%%%%%%%%%%%%%%%%%%%%%%%%%%%%%%%%%
\noindent{\textbf{Phase Diagram---}}Fig.~\ref{fig:tie} 
shows the tie lines computed on the $(\hat{\sigma}_{xy}\!-\!u)$ and
$(\widehat{\dot{\gamma}}\!-\!u)$ planes according to the procedure
outlined in Section~III. Several features should be noted. Flow
induces nematic behavior in what, in equilibrium, would be an
isotropic phase. The tie lines are horizontal in the
$(\hat{\sigma}_{xy}\!-\!u)$ plane, since phases coexist at a
prescribed stress; and have a positive slope in the
$(\widehat{\dot{\gamma}}\!-\!u)$ plane because the more concentrated
nematic phase flows faster. There is a critical point at sufficiently
strong stress, whose existence is expected since the flow-aligning
nematic and paranematic states have the same symmetry
($\boldsymbol{Q}$ is biaxial) and their major axes in the shear plane.

More interesting is the changing slope of the tie lines. For weak
stresses the equilibrium system is slightly perturbed and the tie
lines are almost horizontal. For high stresses the tie lines become
more vertical and the composition difference between the phases
decreases. The slope of the tie lines determines the shape of the mean
stress--strain-rate relation $\bar{\sigma}_{xy}(\bar{\dot{\gamma}})$
that would be measured in steady state experiments.

%%%%%%%%%%%%%%%%%%%%%%%%%%%%%%%%%%%%%%%%%%%%%%%%%%%%%%%%%%%%%%%%%%%
 {\begin{figure}
 \par\columnwidth20.5pc
 \hsize\columnwidth\global\linewidth\columnwidth
 \displaywidth\columnwidth
 \epsfxsize=3.5truein
 \centerline{\epsfbox[160 230 485 660]{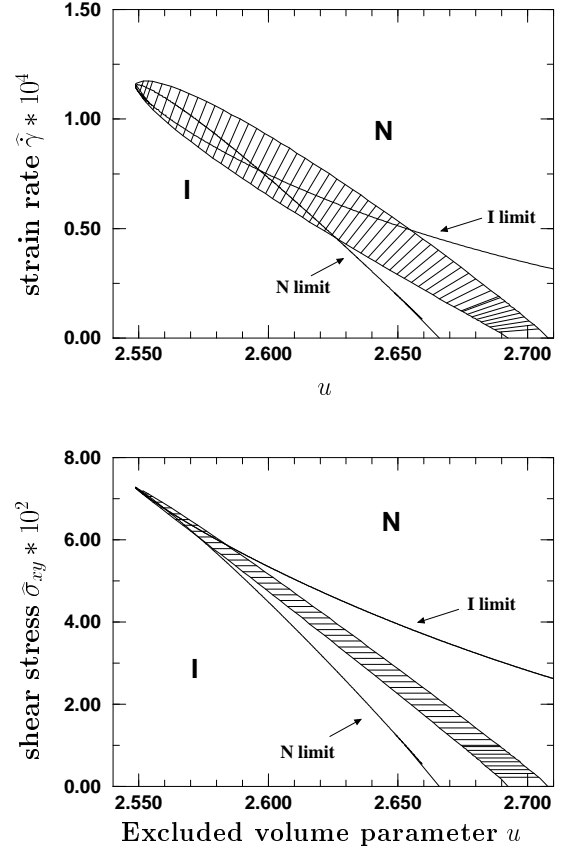}}
 \caption{
   Phase diagram in the $(\hat{\sigma}_{xy}\!-\!u)$ (a) and
   $(\widehat{\dot{\gamma}}\!-\!u)$ (b) planes for $L=5.0,
   \lambda=1.0$, along with the limits of stability of \textsf{I} and
   \textsf{N} phases}
 \label{fig:tie}
 \end{figure}}
%%%%%%%%%%%%%%%%%%%%%%%%%%%%%%%%%%%%%%%%%%%%%%%%%%%%%%%%%%%%%%%%%%%

\end{multicols}\widetext
%%%%%%%%%%%%%%%%%%%%%%%%%%%%%%%%%%%%%%%%%%%%%%%%%%%%%%%%%%%%%%%%%%% 
 {\begin{figure}
 %\displaywidth\columnwidth
 \epsfxsize=\displaywidth
 \centerline{\epsfbox[100 220 1400 635]{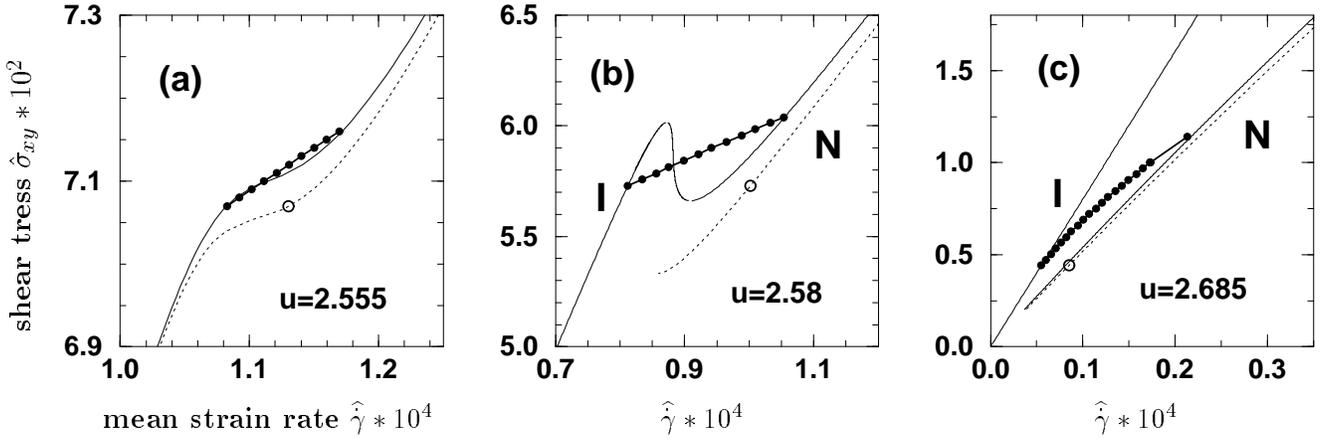}}
 \caption{Mean stress--strain-rate curves for coexistence at common
   stress, for $L=5.0$ and $\lambda=1.0$. The solid lines denotes
   \textsf{I} and \textsf{N} branches; the dotted line in each figure
   denotes the stable \textsf{N} branch with which the \textsf{I} state
   coexists at the low strain rate boundary of the coexistence region,
   at a strain rate marked by an open circle $\boldsymbol{\circ}$.  The
   solid circles $\bullet$ and thick solid line denote the stress that
   would be measured in the banded regime.  Phase coexistence occurs
   between phases of \textit{different} compositions than the mean
   compositions ($u=2.555, 2.58, 2.685$). The unstable portion of the
   homogeneous flow curve is shown in (a) and (b), but not (c). Note
   that the plateaus in the two-phase regions in (a) and (b) rather
   obviously do \emph{not} satisfy an equal area construction with the
   underlying constitutive curve at the mean composition.}
 \label{fig:stress-strainbar0}
 \end{figure}}
%%%%%%%%%%%%%%%%%%%%%%%%%%%%%%%%%%%%%%%%%%%%%%%%%%%%%%%%%%%%%%%%%%%
\begin{multicols}{2} \narrowtext
%%%%%%%%%%%%%%%%%%%%%%%%%%%%%%%%%%%%%%%%%%%%%%%%%%%%%%%%%%%%%%%%%%% 
\noindent{\textbf{Mean Constitutive Relations---}}Consider a 
composition in the range where phase-separation occurs. For small
applied stress $\bar{\sigma}_{xy}(\bar{\dot{\gamma}})$ varies smoothly
until the two-phase region is reached. At this stress, a tiny band of
high strain rate strongly-aligned nematic material appears, with
volume fraction determined by the lever rule, Eq.~(\ref{eq:lev1}). The
mean strain rate $\bar{\dot{\gamma}}$ is determined by the lever rule,
Eq.~(\ref{eq:lev2}), and the measured constitutive relation
$\bar{\sigma}_{xy}(\bar{\dot{\gamma}})$ is non-analytic at this point
(see Fig.~\ref{fig:stress-strainbar0}). As the stress is increased
further, the system traverses the two-phase region by jumping from tie
line to tie line. Each successive tie line has a higher stress, a
higher mean strain rate, and a steadily increasing volume fraction of
nematic phase.  The compositions of both coexisting phases change
steadily through the two-phase region.

The constitutive relation $\bar{\sigma}_{xy}(\bar{\dot{\gamma}})$
through the two phase region is determined by the spacing and splay of
the tie lines. For mean compositions $\bar{\phi}$ close to the
equilibrium isotropic-nematic transition
(Fig.~\ref{fig:stress-strainbar0}c) the tie lines in the
$(\widehat{\dot{\gamma}}\!-\!u)$ plane are fairly flat, so that the
stress $\sigma_{xy}$ changes significantly through the two-phase
region; and the `plateau' has definite curvature, reflecting the
initial splay of the tie lines.  For slightly lower mean compositions
(Fig.~\ref{fig:stress-strainbar0}b) the `plateau' is straighter and
flatter, as can be seen in (Fig.~\ref{fig:stress-strainbar}), because
the lines are more vertical in the $(\widehat{\dot{\gamma}}\!-\!u)$
plane.  Finally, for compositions near the critical point the plateau
is flatter still but, more interestingly, phase coexistence occurs in
a region where the stress-strain curve at the mean composition is no
longer non-monotonic (Fig.~\ref{fig:stress-strainbar0}a)! This is
because stability in a two-phase system is also determined by the
stability with respect to composition variations. In fact, the local
chemical potential $\mu(u)$ has negative slope and is unstable on a
segment of this curve.  The tie line construction is a graphical
expression of the explanation proposed by Schmitt {\sl et al.\/}
\cite{schmitt95}, who attributed a sloped plateau to
composition-dependence of the stress-strain constitutive relation.
The general relation is given by
\begin{equation}
  \frac{\partial\sigma}{\partial\bar{\dot{\gamma}}} = \left[
    \frac{\zeta}{\eta_I} + \frac{1-\zeta}{\eta_N} -
    m(\sigma)\left\{\frac{1-\zeta}{\dot{\gamma}'_N\eta_N} +
      \frac{\zeta}{\dot{\gamma}'_I\eta_I}\right\} 
\right]^{-1},
\end{equation}
where $m(\sigma)$ is the slope of tie line with stress value $\sigma$,
the lines $\{\sigma_I(\phi),\sigma_N(\phi), \dot{\gamma}_I(\phi),
\dot{\gamma}_N(\phi)\}$ bound the phase coexistence domains in the
$\sigma\!-\!\phi$ and $\dot{\gamma}\!-\!\phi$ planes;
$\eta_{k}=\partial\sigma_{k}/\partial{\dot{\gamma}}$ is the local
viscosity of the $k$th branch, and
$\dot{\gamma}'_{k}=\partial\dot{\gamma}_{k}/\partial\phi$.

%%%%%%%%%%%%%%%%%%%%%%%%%%%%%%%%%%%%%%%%%%%%%%%%%%%%%%%%%%%%%%%%%%% 
 {\begin{figure}%\narrowtext
 \par\columnwidth20.5pc
 \hsize\columnwidth\global\linewidth\columnwidth
 \displaywidth\columnwidth \epsfxsize=3.5truein \centerline{\epsfbox[60
   210 660 620]{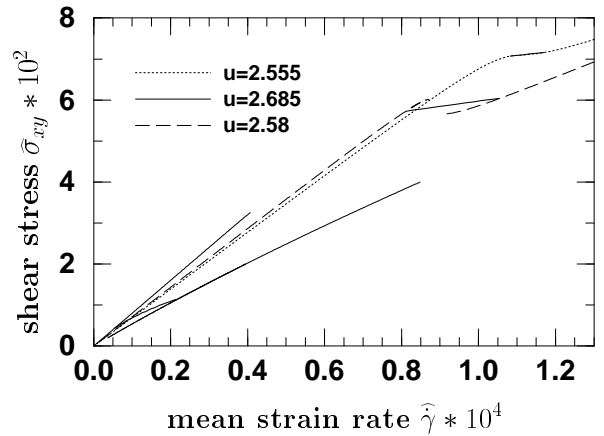}}
 \caption{$\hat{\sigma}_{xy}$ vs. $\widehat{{\dot{\gamma}}}$ 
   for common stress coexistence for $L=5.0$ and $\lambda=1.0$. The
   solid lines connecting the high and low strain rate branches at each
   composition denote the composite flow behavior at coexistence.}
 \label{fig:stress-strainbar}
 \end{figure}}
%%%%%%%%%%%%%%%%%%%%%%%%%%%%%%%%%%%%%%%%%%%%%%%%%%%%%%%%%%%%%%%%%%%
\noindent{\textbf{Measurements at controlled stress or controlled
    strain rate---}}Although these calculations are for phase
separation at a common stress, one may perform experiments at either
controlled stress or strain rate. All three composite curves in
Fig.~\ref{fig:stress-strainbar0} have similar shapes, so we expect the
same qualitative behavior for all compositions. Controlled strain rate
experiments should follow the homogenous flow curves, except for
strain rates in the coexistence regime. Here we expect the steady
state to eventually be the banded state. This should presumably occur
by a nucleation event after some time, for start-up strain rates less
than the \textsf{I} limit shown in Fig.~\ref{fig:tie}a; and occur
immediately for imposed strain rates beyond this stability limit.
Conversely, upon decreasing the strain rate from the nematic phase we
expect nucleated behavior for strain rates larger than the \textsf{N}
limit, and instability for smaller strain rates. In the metastable
regime we expect the flow curve to follow the underlying homogeneous
constitutive curve for the given composition, until the nucleation
event occurs.  Interestingly, there is a small region (inside the loop
in Fig.~\ref{fig:tie}a) where the system is unstable when brought, at
controlled strain rate, into this region from either the \textsf{I} or
\textsf{N} states. This corresponds to constitutive curves with the
multi-valued behavior of curve \textsf{b} in Fig.~\ref{fig:constit}.

Controlled stress experiments should exhibit similar behavior.
Consider Fig.~\ref{fig:stress-strainbar0}b. For initial applied
stresses larger than the minimum coexistence stress and less than the
\textsf{I} limit of stability in Fig.~\ref{fig:tie}b, we expect the
system to follow the homogenous flow curve until a nucleation event
occurs.  After nucleation the strain rate should increase, until
either the proper plateau strain rate or the high strain rate nematic
state is reached, depending on the magnitude of the stress. For
stresses larger than the limit of stability we expect the system to
become immediately unstable to either a banded flow or a homogeneous
nematic phase, depending on the magnitude of the stress.

\noindent{\textbf{Metastability:~Experiments---}}Experiments on
wormlike micelles \cite{berret94a,berret94b,grand97} have found
constitutive curves analogous to those in, say,
Fig.~\ref{fig:stress-strainbar0}. In these experiments the plateau
appears to be the stable states, while the portion of the constitutive
curve (a `spine') which extends to stresses above the onset of the
stress plateau appears to be a metastable branch on which the system
may remain for a finite period of time under controlled stress or
strain rate conditions.  Refs.~\cite{berret94b,grand97,Berr97}
conducted controlled strain-rate experiments and found that the system
follows the composite curves (without `spines' that extend above the
onset of the stress plateau) in Fig.~\ref{fig:stress-strainbar0}, if
care is taken to reach steady state. In these systems the plateaus
were nearly flat, suggesting a very slight dependence of the flow
behavior on composition.  For controlled strain rate quenches into
what corresponds to the two-phase region of Fig.~\ref{fig:tie}a, the
system took some time to develop shear bands and phase separate.  This
relaxation or `nucleation' time decreased as the mean-strain rate was
increased \cite{grand97}.  It is not clear that they reached a limit
of stability (which would be analogous to the \textsf{I} limit in
Fig.~\ref{fig:tie}).  The relaxation times were of order $60-600\,{\rm
  s}$, depending on temperature, mean composition, and mean strain
rate.  We emphasize that these experiments were on micellar solutions,
which probably do not show an isotropic-nematic transition, but still
display the same qualitative stress--strain-rate relationship as curve
\textsf{b} in Fig.~\ref{fig:constit}.

Ref.~\cite{Call+96} revealed different stress plateaus upon
controlling either the strain rate or the shear stress (see Fig.~7 of
Ref.~\cite{Call+96}) in cone-and-plate flow.  In controlled stress
experiments the stress plateau occurred at a stress of order $1.5$
times the stress plateau observed under controlled strain rate
conditions. Moreover, the flow curve under controlled stress
conditions exhibited a stress maximum and then a decrease in stress to
an approximate flat plateau. One explanation for the high stress
plateau under controlled stress conditions could be that the `spine'
never nucleated under controlled stress conditions, and the system
smoothly transformed to the high strain rate phase.  However, we do
not have an explanation for the decrease and subsequent plateau in
stress under applied strain rate conditions.

In other experiments, controlled stress experiments revealed two kinds
of metastable behavior \cite{grand97}. For
$\sigma_p<\sigma<\sigma_{jump}$, where $\sigma_p$ is the minimum
stress for the onset of banding in controlled strain rate experiments,
the system maintained a strain rate on the `metastable' branch for
indefinite times (measured times were up to $10^4\,\hbox{s}$). For
$\sigma>\sigma_{jump}$ the system accelerated, after of order
$10^3\,\hbox{s}$, and left the rheometer.  For these systems it is not
clear whether a stable high shear branch exists. An explanation for
$\sigma_{jump}$ is lacking.  Evidently the nucleation processes
governing metastability at controlled stress and controlled strain
rate are different. Clearly we need more experiments and theory about
the nature of nucleation and metastability in controlled stress vs.
controlled strain rate experiments.

\noindent{\textbf{Polydispersity---}}Fig.~\ref{fig:tieL} shows the
effect of rod aspect ratio $L$ on the phase diagram. A smaller rod
aspect ratio couples more weakly to the flow, requiring a slightly
larger strain rate to induce a transition to the nematic phase
(Fig.~\ref{fig:tieL}a). The resulting stress is slightly smaller
because, when the system enters the two phase region the stress is
largely determined by that of the paranematic branch, which decreases
with increasing $L$ (Fig.~\ref{fig:tieL}b). Although the equilibrium
phase boundaries are close (see Eqs.~\ref{eq:INa}-\ref{eq:INb}), the
deviation is amplified considerably by applying flow.  This suggests
that flow enhances the natural tendency of length polydispersity to
widen biphasic regimes.
%%%%%%%%%%%%%%%%%%%%%%%%%%%%%%%%%%%%%%%%%%%%%%%%%%%%%%%%%%%%%%%%%%%
 {\begin{figure}
 \par\columnwidth20.5pc
 \hsize\columnwidth\global\linewidth\columnwidth
 \displaywidth\columnwidth
 \epsfxsize=3.5truein
 \centerline{\epsfbox[140 230 500 650]{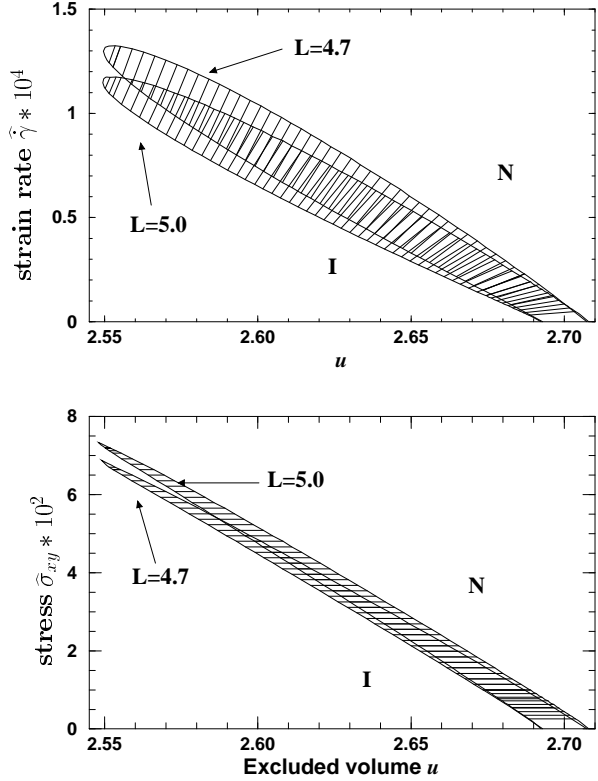}}
 \caption{
   Phase diagrams for $L=5.0$ and $L=4.7$ at common stress, for
   $\lambda=1.0$.}
 \label{fig:tieL}
 \end{figure}}
%%%%%%%%%%%%%%%%%%%%%%%%%%%%%%%%%%%%%%%%%%%%%%%%%%%%%%%%%%%%%%%%%%%

%%%%%%%%%%%%%%%%%%%%%%%%%%%%%%%%%%%%%%%%%%%%%%%%%%%%%%%%%%%%%%%%%%%%%%
\subsection{Paranematic--Log rolling coexistence ({\sf I-L})}
%%%%%%%%%%%%%%%%%%%%%%%%%%%%%%%%%%%%%%%%%%%%%%%%%%%%%%%%%%%%%%%%%%%%

Fig.~\ref{fig:tieIL} shows the phase diagram calculated for
coexistence between paranematic ({\sf I}) and log rolling ({\sf L})
states. As with \textsf{I-N} coexistence, the zero shear limit
corresponds to the equilibrium biphasic region. However, for non-zero
stress the biphasic region shifts in the direction of higher
concentration. This is reasonable, since the stability limit of the
{\sf L} phase shifts to higher concentrations with increasing stress
(Fig.~\ref{fig:stab}).  Note also that, since the {\sf I} and {\sf L}
phases have major axes of alignment in orthogonal directions, there is
no critical point.  Instead, the window of phase coexistence ends when
the {\sf I} phase becomes unstable to the {\sf N} phase.

We have also computed phase coexistence between {\sf N} and {\sf L}
phases. This occurs at much higher compositions ($u> u_{\ast} \agt
3.0$) and has a narrow width in composition due to the very slight
difference in viscosities of the two phases. Unfortunately, we cannot
resolve this coexistence regime accurately within the numerical
precision of our calculations and do not present these results here.

The existence of two possible phase diagrams for common stress phase
separation raises an interesting question. Can one observe {\sf I-L}
coexistence? Notice that {\sf I-L} coexistence can only occur for
samples prepared at concentrations at or above that necessary for
equilibrium phase separation. One could prepare a phase separated
isotropic-nematic mixture and, by wall preparation, field alignment,
sedimentation, or other techniques, separate the phases into two
macroscopic domains with the nematic phase in the log-rolling
geometry.

%%%%%%%%%%%%%%%%%%%%%%%%%%%%%%%%%%%%%%%%%%%%%%%%%%%%%%%%%%%%%%%%%%%
 {\begin{figure}
 \par\columnwidth20.5pc
 \hsize\columnwidth\global\linewidth\columnwidth
 \displaywidth\columnwidth
 \epsfxsize=3.5truein
 \centerline{\epsfbox[140 245 500 645]{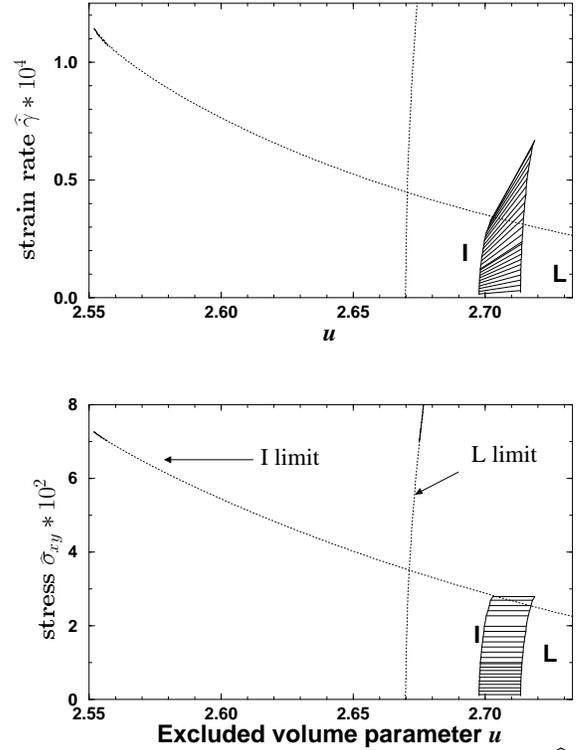}}
 \caption{
   Phase diagram in the $(\hat{\sigma}_{xy}\!-\!u)$ and
   $(\widehat{\dot{\gamma}}\!-\!u)$ planes for paranematic--log-rolling
   coexistence, for $L=5.0$ and $\lambda=1.0$. The dotted lines are the
   limits of stability of the {\sf I} and {\sf L} phases (see
   Fig.~\protect{\ref{fig:stab}}).}
 \label{fig:tieIL}
 \end{figure}}
%%%%%%%%%%%%%%%%%%%%%%%%%%%%%%%%%%%%%%%%%%%%%%%%%%%%%%%%%%%%%%%%%%%

%%%%%%%%%%%%%%%%%%%%%%%%%%%%%%%%%%%%%%%%%%%%%%%%%%%%%%%%%%%%%%%%%%%
 {\begin{figure}
 \par\columnwidth20.5pc
 \hsize\columnwidth\global\linewidth\columnwidth
 \displaywidth\columnwidth \epsfxsize=3.5truein \centerline{\epsfbox[70
   220 740 650]{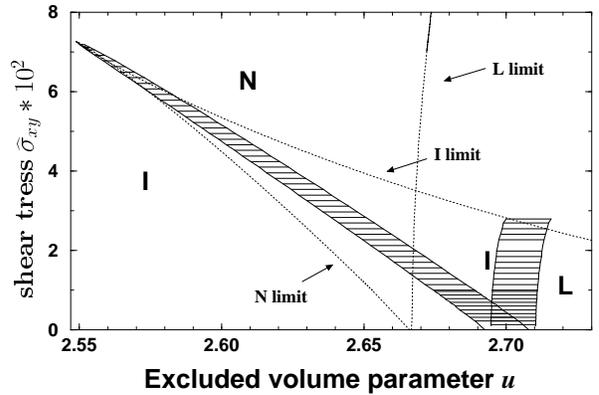}}
 \caption{
   Composite phase diagrams for {\sf I-L} and {\sf I-N} coexistence at
   common stress for $L=5.0$ and $\lambda=1.0$. We stress that this
   represents \emph{two} overlayed phase diagrams, and not a single
   phase diagram. For example, there is \emph{no} triple point implied
   by the intersection of the \textsf{I-N} and \textsf{I-L} phase
   diagrams.}
 \label{fig:tieboth}
 \end{figure}}
%%%%%%%%%%%%%%%%%%%%%%%%%%%%%%%%%%%%%%%%%%%%%%%%%%%%%%%%%%%%%%%%%%%

Upon applying shear, the system could then maintain coexistence and
move through the {\sf I-L} two-phase region.  However, under
controlled strain rate conditions, the {\sf I} material could decay
into {\sf I-N} coexistence (see Fig.~\ref{fig:tieboth}). The resulting
{\sf I-N} coexistence occurs would quickly destabilize the entire {\sf
  I-L} structure.  Therefore the three-band structure {\sf N-I-L} will
not be present in this model, and it is probable that \textsf{I-L}
coexistence could only exist under flow as a metastable state. Similar
conclusions may be drawn by examining the phase diagrams in
field-variable space, $\mu-\sigma_{xy}$, as in Fig.~\ref{fig:mu}a. In
this case the chemical potential of the \textsf{I} phase, at
\textsf{I-L} coexistence, is within the \textsf{N} region of the phase
diagram for \textsf{I-N} coexistence, indicating a (possibly
metastable) instability with respect to \textsf{I-N} phase separation.
Moreover, the chemical potentials of the three phases are never the
same, except at rest where the \textsf{L} and \textsf{N} states are
identical apart from the rod orientations.

%%%%%%%%%%%%%%%%%%%%%%%%%%%%%%%%%%%%%%%%%%%%%%%%%%%%%%%%%%%%%%%%%%%%%%%%%%%%
\section{Common Strain Rate Coexistence}
%%%%%%%%%%%%%%%%%%%%%%%%%%%%%%%%%%%%%%%%%%%%%%%%%%%%%%%%%%%%%%%%%%%%%%%%%%%%
For coexistence at common strain rate the interface lies in the
velocity--velocity-gradient plane, and inhomogeneities are in the
${\rm\bf\hat{z}}$ direction (see Fig.~\ref{fig:couette}). The stress
balance condition at the interface is
$\boldsymbol{\sigma}\!\cdot\!{\rm\bf\hat{z}}$ uniform.  As before,
$\sigma_{zz}$ is taken care of by the pressure while $\sigma_{yz}$ and
$\sigma_{xz}$ are zero by symmetry (and because there are no stable
$q_3$ components in the order parameter tensor).  With bands in the
${\rm\bf\hat{z}}$ direction, the strain rate in each band is set by
the relative velocity of the two plates (or cylinders, in a Couette
device), and the shear stresses differ.  The mean applied stress
$\bar{\sigma}_{xy}$ is the area average of the stress applied to each
band.  The coexisting phases have shear stresses and compositions
partitioned according to
\begin{eqnarray} 
\bar{\phi} &=& \zeta \phi_{\scriptscriptstyle
I} + (1-\zeta) \phi_{\scriptscriptstyle II} \label{eq:lev1s}\\
\bar{\sigma}_{xy} &=& \zeta \sigma_{xy}^{\scriptscriptstyle I} + (1-\zeta)
\sigma_{xy}^{\scriptscriptstyle II}, \label{eq:lev2s} 
\end{eqnarray}
where $\bar{\sigma}_{xy}$ is the mean shear stress. The interfacial
equations to solve are Eqs.~(\ref{eq:Q}),(\ref{eq:sigma}), and
~(\ref{eq:mu}).

\noindent{\textbf{Phase Diagram---}}Common strain rate
\textsf{I-N} phase coexistence is shown in Fig.~\ref{fig:ties}. In
this case the tie lines are horizontal in the
$(\widehat{\dot{\gamma}}\!-\!u)$ plane. They have a negative slope in
the $(\hat{\sigma}_{xy}\!-\!u)$ plane because the paranematic {\sf I}
phase coexists with a denser and less viscous flow-aligning \textsf{N}
phase.  As with phase separation at common stress, there is a (very
small) loop in the limits of stability in the control variable plane
($\dot{\gamma}\!-\!u$) within which there are no stable homogeneous
states. The careful reader will note that the limits of stability at a
given stress (Fig.~\ref{fig:tie}) are different from the limits of
stability at a given strain rate. This is physically correct, and will 
be discussed below in Sec.~\ref{sec:spinodals}.

%%%%%%%%%%%%%%%%%%%%%%%%%%%%%%%%%%%%%%%%%%%%%%%%%%%%%%%%%%%%%%%%%%%
 {\begin{figure}
 \par\columnwidth20.5pc
 \hsize\columnwidth\global\linewidth\columnwidth
 \displaywidth\columnwidth
 \epsfxsize=3.0truein
 \centerline{\epsfbox[170 230 480 650]{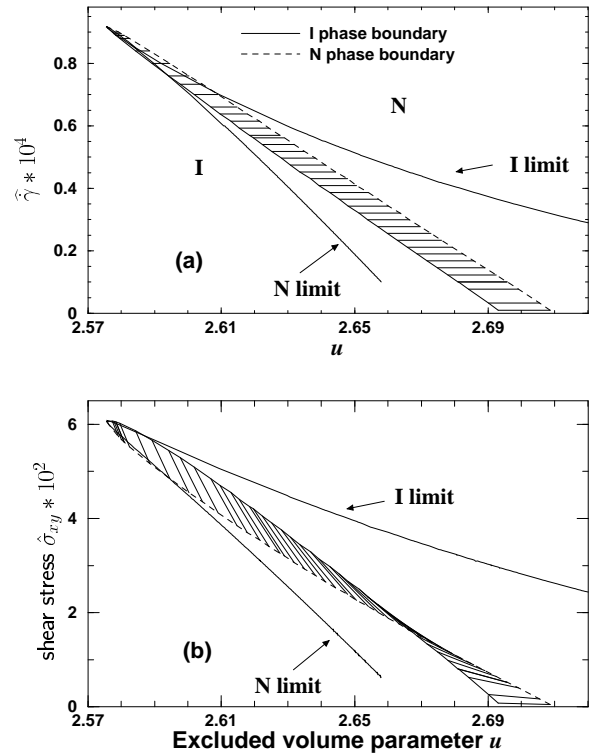}}
 \caption{Common strain rate phase diagram in the
   $(\widehat{\dot{\gamma}}\!-\!u)$ (a) and $(\hat{\sigma}_{xy}\!-\!u)$
   (b) planes, for $L=5.0$ and $\lambda=1.0$. Also shown are the limits
   of stability of the \textsf{I} and \textsf{N} phases (calculated for
   a given imposed strain rate, in contrast to Figures \ref{fig:stab},
   \ref{fig:tie}, \ref{fig:tieIL}, and \ref{fig:tieboth}, in which the
   stability was calculated for an imposed stress.)}
 \label{fig:ties}
 \end{figure}}
%%%%%%%%%%%%%%%%%%%%%%%%%%%%%%%%%%%%%%%%%%%%%%%%%%%%%%%%%%%%%%%%%%%
There is an interesting crossover visible in the
$(\hat{\sigma}_{xy}\!-\!u)$ plane.  For higher mean compositions the
fluid has a higher stress in its high strain rate one-phase region
than in its low strain rate one-phase region; that is, respectively
above and below the biphasic region in the Fig.~\ref{fig:ties}a.
Conversely, for low enough compositions $u\alt 2.67$, the stress in
the high strain rate region immediately outside the biphasic regime is
actually {\sl less\/} than the stress just before the system enters
the biphasic region, as can be seen by the crossing of the solid and
dashed phase boundaries in Fig.~\ref{fig:ties}.

This crossover is straightforward to understand. Since phase
separation occurs at a given strain rate, and the stress of the {\sf
  N} branch at a given composition and strain rate is always less than
that of the corresponding {\sf I} branch, we expect a decrease in the
stress upon leaving the biphasic regime in cases where the coupling to
composition is less important. We saw in the analysis at common stress
that composition effects are less important (for {\sf I-N}
coexistence) at lower compositions and high strain rates, where the
tie lines are more vertical. We expect this near the critical point
where the two phases become more and more similar.  More generally, we
expect this behavior in situations where phase separation occurs at a
common strain rate into a shear-thinning state with only slight
changes in composition. In the more concentrated regime, the
coexistence plateau traverses a wider range of concentrations and
strain rates, and emerges into the pure \textsf{N} phase with a higher
stress (the width in strain rate of the phase coexistence regime is
enough to overcome the shear thinning effect of the nematic phase).

\noindent{\textbf{Mean Constitutive Relations---}}
Figs.~\ref{fig:stressstrainbars0}-\ref{fig:stressstrainbars} show the
mean stress--strain-rate relations. As with common stress phase
separation, the shape of the `plateau' as the strain rate is swept
through the two-phase region is not always flat, and depends on the
splay of the tie lines. At higher concentrations the plateau has a
positive slope while, in accord with the crossover in the
$(\hat{\sigma}_{xy}\!-\!u)$ phase diagram, for lower concentrations
the plateau crosses over to negative slope, which usually signifies a
bulk instability. A simple argument, analogous to that for the
stability of a bulk fluid, supports this. However, we note that a
composite negative slope curve was accessed, and apparently found
stable, by Hu \emph{et al.}  \cite{HBP98} under controlled stress
conditions.  The negative slope in Fig.~\ref{fig:stressstrainbars0}a
is likely to be inaccessible under controlled stress conditions, and
the instability argument may apply to controlled strain rate
conditions.  The general relation for the slope in the composite
region is \cite{schmitt95}
\begin{equation}
  \frac{\partial\bar{\sigma}}{\partial{\dot{\gamma}}} = 
    \eta_I\,\zeta + \eta_N\,(1-\zeta) -
    m(\dot{\gamma})\left\{\frac{\eta_N(1-\zeta)}{\sigma'_N} +
      \frac{\eta_I\,\zeta}{\sigma'_I}\right\},
\end{equation}
where $m(\dot{\gamma})$ is the slope of the tie line with strain rate
$\dot{\gamma}$ and $\sigma'_{k}=\partial\sigma_{k}/\partial\phi$.  In
the limit of no concentration difference ($\delta\phi=0$ or
$m(\dot{\gamma})=\infty$), $\sigma(\dot{\gamma})$ is vertical through
the two phase region.

%%%%%%%%%%%%%%%%%%%%%%%%%%%%%%%%%%%%%%%%%%%%%%%%%%%%%%%%%%%%%%%%%%%
\end{multicols}\widetext
%%%%%%%%%%%%%%%%%%%%%%%%%%%%%%%%%%%%%%%%%%%%%%%%%%%%%%%%%%%%%%%%%%% 
 {\begin{figure}
 \displaywidth\columnwidth
 \epsfxsize=\displaywidth
 \centerline{\epsfbox[80 210 1300 600]{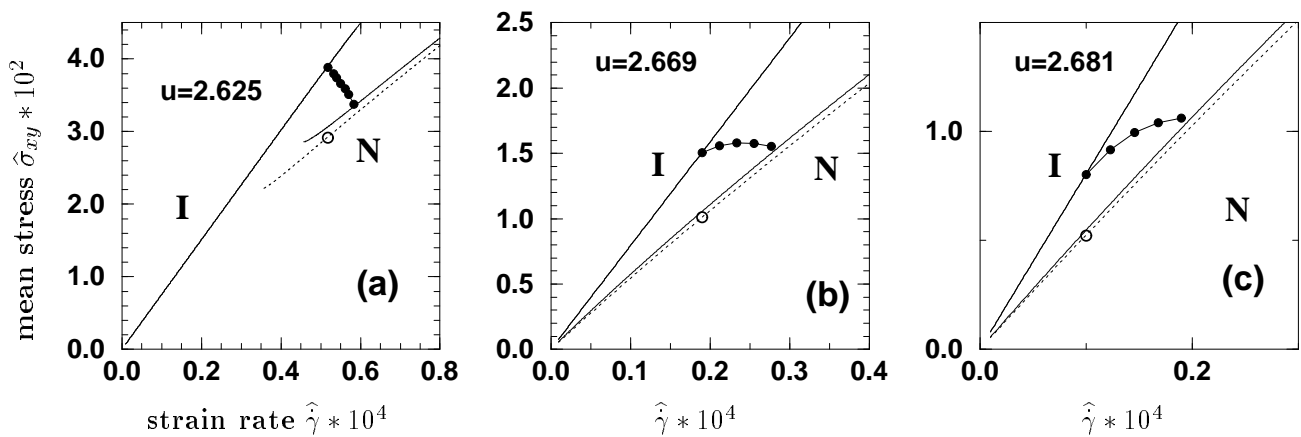}}
 \caption{Mean stress--strain-rate curves for common strain 
   rate coexistence for $L=5.0$ and $\lambda=1.0$.  The solid lines
   denote the stable \textsf{I} and \textsf{N} branches; the dotted
   line in each figure denotes the stable \textsf{N} branch with which
   the \textsf{I} state coexists at the low strain rate boundary of the
   coexistence region, at a strain rate marked by an open circle
   $\boldsymbol{\circ}$.  The solid circles $\bullet$ and thick solid
   line denote the stress that would be measured in the banded regime.
   The filled circles $\bullet$ and thick solid line denotes the stress
   measured under banded conditions. }
 \label{fig:stressstrainbars0}
 \end{figure}}
%%%%%%%%%%%%%%%%%%%%%%%%%%%%%%%%%%%%%%%%%%%%%%%%%%%%%%%%%%%%%%%%%%%
\begin{multicols}{2}

\noindent{\textbf{Measurements at controlled stress or controlled
    strain rate---}}For controlled strain rate measurements we expect
behavior similar to that for phase separation at common stress. For
start-up experiments with mean strain rates larger than the minimum
strain rate for coexistence at a given composition, we expect the
stress to follow the metastable branch until a nucleation event causes
the stress to decrease to the plateau stress. The exception is a
composition such as that in Fig.~\ref{fig:stressstrainbars0}a, for
which the composite flow curve for \textsf{I-N} coexistence may be
mechanically unstable.  Similar results should apply upon decreasing
the strain rate from the shear-induced \textsf{N} phase to below the
\textsf{N} limit. As before, this expectation of a nucleation event is
based on a possibly misguided analogy with equilibrium systems which,
nonetheless, is encouraging given the experiments which see
``nucleation'' type behavior in micelles under flow
\cite{berret94b,grand97,Berr97}.

For controlled stress, the situation is slightly different. For
compositions with mean stress--strain-rate curves of the shape of
Fig.~\ref{fig:stressstrainbars0}c, we expect similar behavior to that
found for common stress phase separation. However, for compositions
that yield curves such as Fig.~\ref{fig:stressstrainbars0}a there is a
window of stresses for which there are \emph{three} possible states:
homogeneous low strain rate and high strain rate branches, and a
banded intermediate branch. We emphasize that we have not determined
the absolute stability of any of these branches.  A possibility is
that the system has hysteretic behavior. For example, in start-up
experiments the system would remain on the \textsf{I} branch until a
certain stress, at which point it would nucleate after some time and
transform to either the high strain rate \textsf{N} branch or
coexistence. We cannot tell which state it might go to, from this
analysis, but it seems likely that it would jump straight to the
\textsf{N} branch.\footnote{If the system jumped from the \textsf{I}
  branch to the coextence branch, increasing the stress further would
  \emph{decrease} the strain rate and return the system to the
  \textsf{I} branch.  Since it originally nucleated \emph{from} the
  \textsf{I} branch, it seems unlikely that the original jump could be
  to the coexisting plateau.}  The same behavior (in reverse) would be
expected upon reducing the stress from the high strain rate \textsf{N}
phase.

%%%%%%%%%%%%%%%%%%%%%%%%%%%%%%%%%%%%%%%%%%%%%%%%%%%%%%%%%%%%%%%%%%% 
 {\begin{figure}%\narrowtext
 \par\columnwidth20.5pc
 \hsize\columnwidth\global\linewidth\columnwidth
 \displaywidth\columnwidth \epsfxsize=3.5truein \centerline{\epsfbox[60
   270 520 570]{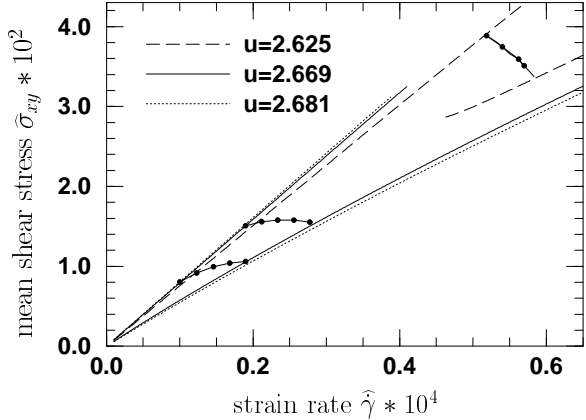}}
 \caption{$\hat{\sigma}_{xy}$ vs. $\widehat{{\dot{\gamma}}}$ 
   for various compositions, for phase separation at common strain rate
   and $L=5$. }
 \label{fig:stressstrainbars}
 \end{figure}}
%%%%%%%%%%%%%%%%%%%%%%%%%%%%%%%%%%%%%%%%%%%%%%%%%%%%%%%%%%%%%%%%%%%

Although there have been anecdotal reports of shear banding in the
common strain rate geometry, there have been very few such results
published. Bonn \emph{et al.} \cite{Bonn+98} have recently reported
results for sheared surfactant onion gels, along with visual
confirmation of bands in the common strain rate geometry. In
controlled strain rate experiments they found constitutive curves
analogous to Fig.~\ref{fig:stressstrainbars0}a
or~\ref{fig:stressstrainbars0}b. In controlled stress experiments they
found hysteretic behavior, with the system flipping between high and
low strain rate branches after some delay time, missing the
coexistence `plateau' regime. However, it is not clear that these were
true steady state results.

Stable `negative-slope' behavior was seen in a shear-thickening
systems which phase separates at common stress
\cite{boltenhagen97a,boltenhagen97b}, under controlled stress
conditions. In this case there was a single (mean) strain rate for a
given applied stress, and the measured constitutive relation had an
\textsf{S} shape rather than the sideways \textsf{S} shape of
Fig.~\ref{fig:stressstrainbars0}.

%%%%%%%%%%%%%%%%%%%%%%%%%%%%%%%%%%%%%%%%%%%%%%%%%%%%%%%%%%%%%%%%%%%%%%%%%%%%
\section{Discussion}\label{sec:discussion}
%%%%%%%%%%%%%%%%%%%%%%%%%%%%%%%%%%%%%%%%%%%%%%%%%%%%%%%%%%%%%%%%%%%%%%%%%%%%
\subsection{Dependence on gradient terms}
%%%%%%%%%%%%%%%%%%%%%%%%%%%%%%%%%%%%%%%%%%%%%
Gradient terms appear in all equations of motion for $K\neq0$ and for
any $g$, so to avoid unphysical equations without gradients (which
cannot resolve interfaces) we must have $\lambda\sim g/K<\infty$.  In
the case of $K=0$ and finite $g$ the $\boldsymbol{Q}$ equation of
motion has no explicit gradient terms and hence can, in principle,
support discontinuous solutions. The $\phi$ equation has gradients in
this case, arising from the term $g\left(\nabla\phi\right)^2$ in the
free energy density, Eq.~(\ref{eq:free}), so the system will
eventually reach a state with smooth solutions in both $\phi$ and
$\boldsymbol{Q}$.  Conversely, for $g=0$ there are gradient terms in
both the $\boldsymbol{Q}$ and $\phi$ dynamics, with the latter arising
from the term $\phi\left(\nabla\boldsymbol{Q}\right)^2$ in the free
energy density Eq.~(\ref{eq:free}).
  
Phase boundaries for $\hat{\sigma}_{xy}=0.01, 0.03$ are shown in
Fig.~\ref{fig:acc}. For $\lambda\in(0.0-30.0)$ the phase boundaries
are the same, within the precision of our numerical calculations,
while there is a distinct difference for $\lambda=\infty$. We have
discretized the system on a mesh of 125 points, and the range of
elastic constants is such that the width of the interface is at least
20 mesh points; large enough for smooth behavior and much smaller than
the system size.

%%%%%%%%%%%%%%%%%%%%%%%%%%%%%%%%%%%%%%%%%%%%%
 {\begin{figure}
 \par\columnwidth20.5pc
 \hsize\columnwidth\global\linewidth\columnwidth
 \displaywidth\columnwidth
 \epsfxsize=3.5truein
 \centerline{\epsfbox[50 200 750 650]{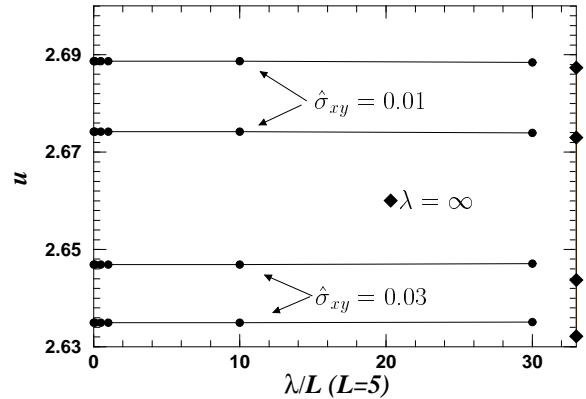}}
 \caption{\textsf{I-N} phase boundaries for common stress
   phase separation as a function of $\lambda/L$, for $L=5.0$. The
   diamonds $\blacklozenge$ are for $\lambda=\infty (K=0,g=1)$.}
 \label{fig:acc}
 \end{figure}}
%%%%%%%%%%%%%%%%%%%%%%%%%%%%%%%%%%%%%%%%%%%%

We cannot rule out the possibility that changes in $\lambda$ shift the
phase boundaries by small amounts below our accuracy, which is of
order 0.1\% in $u$, but the apparent independence of the phase
boundaries on $\lambda$ is curious. One might be tempted to generalize
and suggest that, for finite $\lambda$, there exists a selection
criterion which involves only the homogeneous equations of motion,
rather than requiring the inhomogenous terms as in the interface
construction. An interface construction may also be used to determine
equilibrium phase boundaries, in which case a stationary interface is
equivalent to minimizing a free energy and the (relaxational)
dynamical equations derived from a variational principle
\cite{olmsted92}. In the case of a van der Waals fluid this reproduces
the Maxwell construction.

A steady state equation for a single variable $\psi$ with homogeneous
and inhomogeneous terms of the form
\begin{equation}
  \label{eq:4}
  \sigma_0 = f_{\mit hom}(\psi) + f_{\mit inh}(\partial\psi/\partial y)
\end{equation}
can be integrated to yield a solvability condition for $\sigma_0$,
which is equivalent to the stable interface method.  In equilibrium
$f_{\mit inh}$ integrates exactly without an integrating factor, since
it typically arises from a variation of a free energy functional with
respect to $\psi$, and the result $\sigma_0$ (corresponding to the
pressure in the van der Waals fluid) depends only on $f_{\mit hom}$.
Out of equilibrium, integration is not so simple, and the solvability
condition depends, generally, on the form of the gradient terms
\cite{pomeau86,jsplanar}.

In the multivariable case considered here the steady state conditions
for the order parameter and composition are coupled differential
equations which are not integrable in shear flow. This is because of
the terms $\boldsymbol{\kappa}\!\cdot\!\boldsymbol{Q} +
\boldsymbol{Q}\!\cdot\!\boldsymbol{\kappa}^{\scriptstyle T}$ in
Eq.~(\ref{eq:convect}) and
$(\nabla^2\boldsymbol{Q})\!\cdot\!\boldsymbol{Q} -
\boldsymbol{Q}\!\cdot\!(\nabla^2\boldsymbol{Q})$ in
Eq.~(\ref{eq:sigma}).  In extensional flow $\boldsymbol{\kappa}$ is
symmetric, so that $\boldsymbol{\kappa}\!\cdot\!\boldsymbol{Q} +
\boldsymbol{Q}\!\cdot\!\boldsymbol{\kappa}^{\scriptstyle T}$
integrates to ${\rm Tr}(Q^2\boldsymbol{\kappa})/2$, while in shear
flow this term can only be integrated by introducing an integral
representation \cite{zwillinger}. Hence a first integral of the steady
state equations cannot be found in shear flow, and it seems unlikely
that a general condition involving only the homogeneous portion of the
steady state equations can determine coexistence.  While we appear to
find, for this set of gradient terms, solvability conditions that are
independent of $\lambda$ for $\lambda<\infty$, the relationship of
this to a variational principle remains unknown.  We have not
exhausted the possible gradient terms. For example, higher order
gradients in the free energy ($(\nabla^2\boldsymbol{Q})^2$,
\emph{etc.}) would yield higher-order differential equations for the
interfacial profile, and other square gradient terms such as
$Q_{\alpha\beta}\nabla_{\alpha}\nabla_{\beta}\phi$ are possible
\cite{liu93}.  Hence, we believe that, for finite $\lambda$,
\textit{the apparent independence of our results on gradient terms
  only applies to the particular (simple) family of gradient terms we
  have chosen}. The structure of the differential equations describing
the steady states may change abruptly for $\lambda=\infty$, for which
a term is lost in the differential equations, leading to a distinctly
different selection criteria and the shifted phase boundary in
Fig.~\ref{fig:acc}. Unfortunately, this particular set of equations is
too complex for this kind of analysis.
For example, in a study of a
simpler constitutive model, one can demonstrate that the selected
stress depends on the detail form of the gradient terms
\cite{jsplanar}.  

Several workers have claimed to find an equal areas construction on
the stress--strain-rate constitutive curve \cite{greco97}.  That is,
the ``plateau'' as the system traverses the two-phase region is said
to describe a path such that the areas above and below the plateau,
enclosed by the plateau line and the underlying constitutive curve,
are the same. This is not true here, as can be seen in
Fig.~\ref{fig:stressstrainbars0}.

%%%%%%%%%%%%%%%%%%%%%%%%%%%%%%%%%%%%%%%%%%%%%%%%%%%%%%%%%%%%%%%
\subsection{Which phase separation is preferred?}\label{which}
%%%%%%%%%%%%%%%%%%%%%%%%%%%%%%%%%%%%%%%%%%%%%%%%%%%%%%%%%%%%%%%

Having calculated \textit{both} common strain rate and common stress
phase separation for the same system, and noticing from
Figs.~\ref{fig:tieboth} and~\ref{fig:ties} that there are compositions
and shear conditions which lie inside the two-phase regions of all
three calculated phase separations, we must address the question of
which phase separation occurs. We have already argued that we expect
\textsf{I-L} phase-separation at common stress to be metastable with
respect to \textsf{I-N} phase separation at common stress.  What about
the relative stability of \textsf{I-N} phase separation at either
common stress or common strain rate?

With limited one-dimensonal calculations for systems of different
symmetry (annular bands at common stress and stacked disk-like bands
for common strain rate) it is impossible to calculate the stability of
one interface profile with respect to another. Renardy calculated the
stability of common stress coexistence to capillary fluctuations
\cite{renardy}, which is a start; and such a stability analysis has
been performed, in part, on the layer orientation of smectic systems
in flow \cite{goulian95}. However, some insight can be obtained by
examining the ``phase diagrams'' in the chemical potential--field
variable (either stress or strain rate) planes.  The solid lines in
Fig.~\ref{fig:mu} are analogous to lines of phase coexistence in, for
example, the pressure--temperature plane in a simple fluid.

Consider Fig.~\ref{fig:mu}a. Here, $\widehat{\sigma}_{xy}$ and $\mu$
are the proper field variables for phase separation at a common
stress, and the solid lines denote the \textsf{I-L} and \textsf{I-N}
phase boundaries.  The dashed line denotes the range of stresses at
coexistence for common strain rate phase separation
(Fig.~\ref{fig:mu}a), for which stress is a generalized density
variable and strain rate the field variable.

Fig.~\ref{fig:mu}a indicates that, for a system undergoing
\textsf{I-N} at a common strain rate, the chemical potential and
stress for the \textsf{I} phase falls within the single phase
\textsf{N} region of the common \emph{stress} phase diagram. Hence, we
expect this \textit{I} phase to be unstable (or metastable) with
respect to phase separation at common stress. Similarly, the control
parameters (chemical potential and stress) for the \textsf{N} phase
coexisting at a common strain rate lie within the single phase
\textsf{I} region for common stress phase separation, which we also
expect to be unstable (or metastable). Conversely, for a system
coexisting at a common stress the \textsf{I} phase lies within the
single phase \textsf{I} region of the common strain rate phase diagram
(Fig.~\ref{fig:mu}b), and similarly for the \textsf{N} phase.  This
suggests that phase separation at a common strain rate is unstable (or
metastable) with respect to phase separation at common stress, while
phase separation at a common stress is stable.
%%%%%%%%%%%%%%%%%%%%%%%%%%%%%%%%%%%%%%%%%%%%%
 {\begin{figure}
 \par\columnwidth20.5pc
 \hsize\columnwidth\global\linewidth\columnwidth
 \displaywidth\columnwidth
 \epsfxsize=3.5truein
 \centerline{\epsfbox[130 230 500 680]{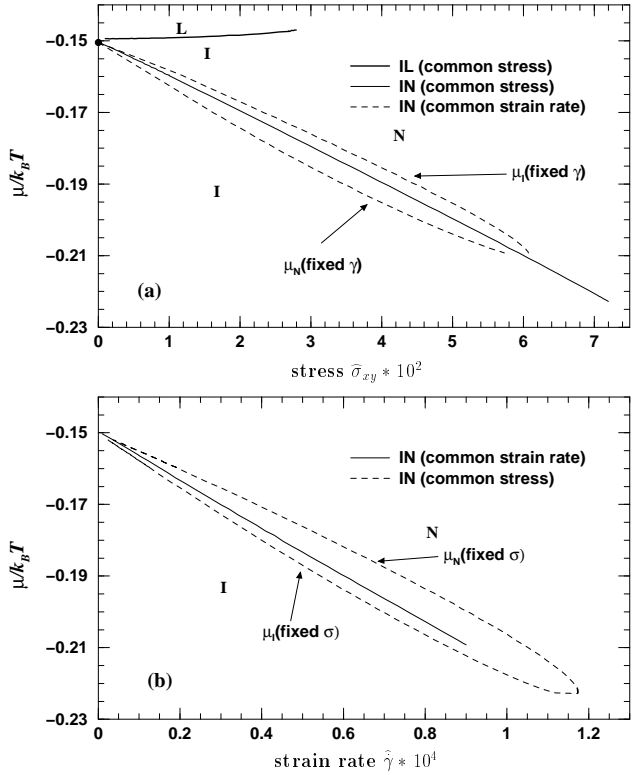}}
 \caption{Phase diagrams in the chemical potential $\mu$ vs. stress plane
   (a) and the $\mu$-strain rate plane (b). The solid lines denote the
   phase boundaries for common stress phase coexistence in the
   $\mu\!-\!\sigma$ plane (a) and for common strain rate coexistence in
   the $\mu\!-\!\dot{\gamma}$ plane (b). The dashed lines denote the
   coexisting stresses for common strain rate phase separation within
   the common stress phase diagram (a); and vice versa in (b).}
 \label{fig:mu}
 \end{figure}}
%%%%%%%%%%%%%%%%%%%%%%%%%%%%%%%%%%%%%%%%%%%%

Note that, ultimately, this selection of phase coexistence geometries
follows from the transition being a shear-thinning transition; for a
shear thickening transition the situation could be reversed. In this
case phase coexistence at a common strain rate and a given $\mu$ would
imply a shear-induced phase (analogous to the \textsf{N} phase) with a
higher stress than the \textsf{I} phase.  If the phase coexistence
line for common stress (strain rate) lay within a loop corresponding
to the stresses (strain rates) for common strain rate (stress)
coexistence, then common strain rate coexistence would be expected to
be stable, by analogy with the isotropic-nematic shear thinning model.
Obviously this argument is delicate. In a fluid where only one phase
coexistence (either common stress or common strain rate) is supported
by the dynamical equations this argument is moot.

%%%%%%%%%%%%%%%%%%%%%%%%%%%%%%%%%%%%%%%%%%%%%%%%%%%%%%%%%%%%%%%%%%%
\end{multicols}\widetext
%%%%%%%%%%%%%%%%%%%%%%%%%%%%%%%%%%%%%%%%%%%%%%%%%%%%%%%%%%%%%%%%%%% 
 {\begin{figure}
 \displaywidth\columnwidth
 \epsfxsize=\displaywidth
 \centerline{\epsfbox[26 102 571 410]{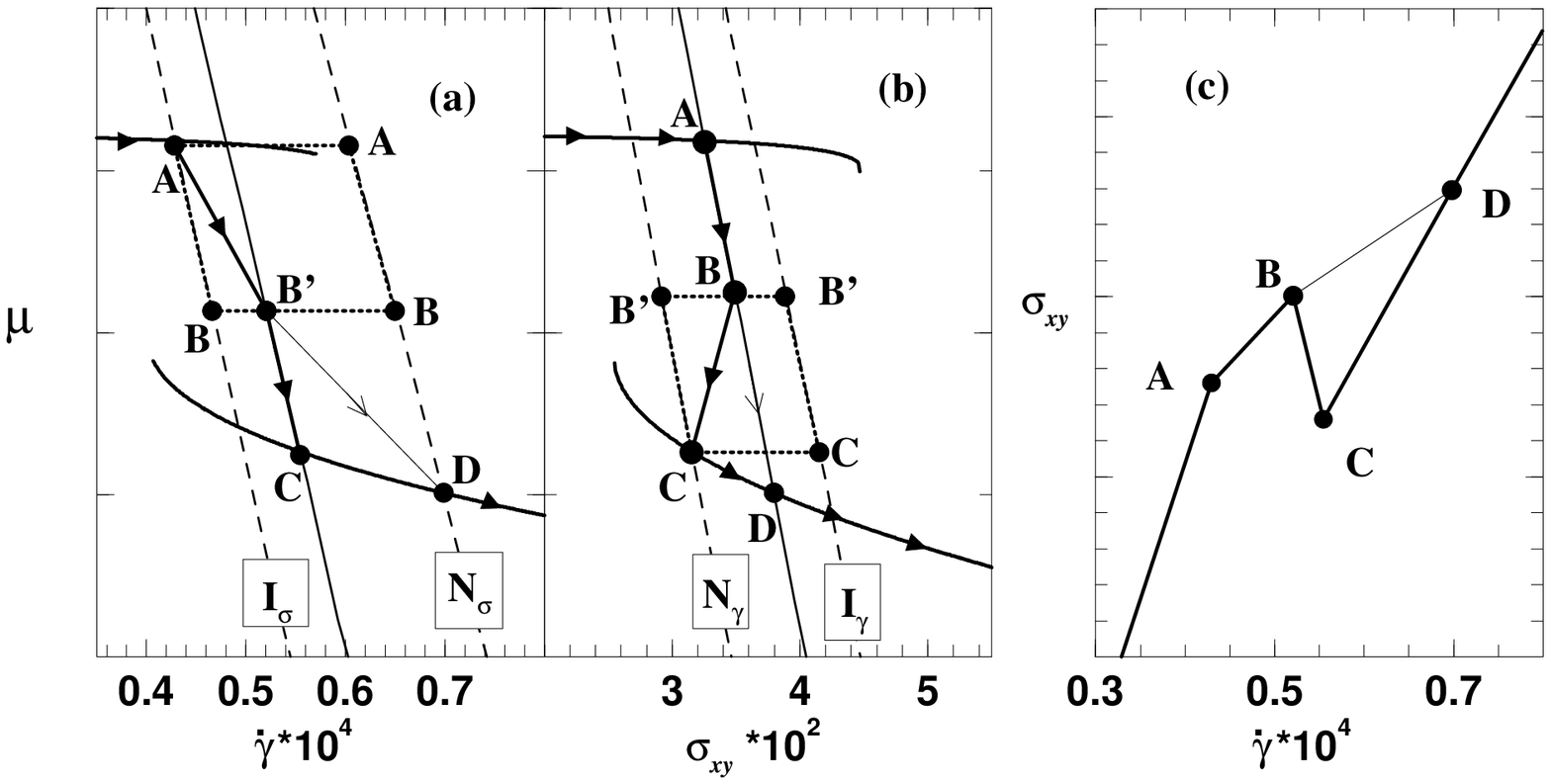}}
\caption{Phase diagrams in the (a) $\mu\!-\!\dot{\gamma}$ and (b)
  $\mu\!-\!\sigma_{xy}$ planes for \textsl{I-N} coexistence (the
  \textsl{N} state is stable for higher strain rate or stress,
  respectively).  The thin vertical solid lines passing through {\tt
    B'-C} and {\tt A-B-D} denote phase coexistence at common strain
  rate and stress in (a) and (b), respectively.  The broken lines
  marked $I_\gamma$ and $N_\gamma$ denote the coexisting states at
  common strain rate in the $\mu\!-\!\sigma_{xy}$ plane (b); while the
  broken lines $I_\sigma$ and $N_\sigma$ denote the coexisting states
  at common stress, in the $\mu\!-\!\dot{\gamma}$ plane. (c) is the
  mean stress vs. strain rate curve.  Shown is a path {\tt A-B-C-D}
  taken under the proposition that the system maintains a global
  minimum in chemical potential. Point {\tt A} is at coexistence in
  the $\mu\!-\!\sigma_{xy}$ plane (b), and hence corresponds to two
  points, on lines $I_{\sigma}$ and $N_{\sigma}$, in the
  $\mu\!-\!\dot{\gamma}$ plane (a) for the two different strain rates
  of the coexisting phases. Similarly, point {\tt C} corresponds to
  coexistence at common strain rate in (a), with the coexisting phases
  at different stresses lying on lines $I_{\gamma}$ and $N_{\gamma}$
  in (b), at the two points {\tt C}.  Points {\tt B} and {\tt B'} are
  coincident in (c), and correspond to a point switching from phase
  separation at common stress ({\tt B}) to phase separation at common
  strain rate ({\tt B'}).  The path {\tt A-B-C} in (c) may be traced
  in (b) by following the upper horizontal arrow until phase
  separation at common stress occurs at {\tt A}, then along the
  segments {\tt A-B} in (b) or {\tt A-B'} in (a) until phase
  separation at common strain rate occurs at {\tt B'}.  From this
  point until {\tt C} the system phase separates along $I_{\gamma}$
  and $N_{\gamma}$, with a mean stress given by the thick diagonal
  solid arrow {\tt B-C} in (b) and the thick segment {\tt B'-C} in
  (a). The system emerges from the two-phase region at {\tt C} on
  $N_{\gamma}$, and continues through {\tt D} on the high strain rate
  branch. }
 \label{fig:resolve}
 \end{figure}}
%%%%%%%%%%%%%%%%%%%%%%%%%%%%%%%%%%%%%%%%%%%%%%%%%%%%%%%%%%%%%%%%%%%
\begin{multicols}{2}
%%%%%%%%%%%%%%%%%%%%%%%%%%%%%%%%%%%%%%%%%%%%%%%%%%%%%%%%%%%%%%%%%%% 

An alternative possibility is presented in Fig.~\ref{fig:resolve} if
one argues that in steady state, among the possible phases which are
compatible with the interface solvability condition, the chemical
potential reaches its absolute minimum.  Consider increasing the
strain rate for a given mean concentration. The thick horizontal
arrows in Fig.~\ref{fig:resolve}a-c denote the $\mu(\sigma_{xy})$ and
$\mu(\dot{\gamma})$ paths for the homogeneous high and low shear rate
states, in the two phase diagrams.  In Fig.~\ref{fig:resolve}a the
path is {\tt A-B'-C-D}, in Fig.~\ref{fig:resolve}b the path is {\tt
  A-B-C-D}, and in Fig.~\ref{fig:resolve}c the path is {\tt
  A-B-C-D}.

This path ensures that the system maintains the minimum chemical
potential for an imposed strain rate. Upon increasing the strain rate
from zero the system remains in the one phase region until {\tt A} is
reached, at which point phase separation at common stress occurs. Note
that {\tt A} spans two points of coexistence in the
$\mu\!-\!\dot{\gamma}$ plane (Fig.~\ref{fig:resolve}a) on the lines
$I_{\sigma}$ and $N_{\sigma}$. Upon further increasing the strain rate
the system continues to phase separate at common stress, following the
segment {\tt A-B} in Fig.~\ref{fig:resolve}b and the two (coexisting)
segments {\tt A-B} in Fig.~\ref{fig:resolve}a. The mean chemical
potential and strain rate follow the diagonal segment {\tt A-B'} in
Fig.~\ref{fig:resolve}a. Upon increasing the \emph{strain rate} above
{\tt B}, the system can continue to maintain its lowest chemical
potential by phase separating with a common strain rate in the two
phases, \emph{i.e.} with shear bands in the vorticity direction.
Hence, the system next follows the path {\tt B'-C} in
Fig.~\ref{fig:resolve}a ($\mu\!-\!\dot{\gamma}$ plane) and the two
coexisting paths {\tt B'-C} in Fig.~\ref{fig:resolve}b
($\mu\!-\!\sigma_{xy}$ plane), with the mean chemical potential and stress
following the diagonal segment {\tt B-C} in Fig.~\ref{fig:resolve}b.
Finally, upon increasing the strain rate above {\tt C} the system
continues along the high strain rate branch. The thin diagonal lines
with arrows, {\tt B'-D} in Fig.~\ref{fig:resolve}a and {\tt B-D} in
Fig.~\ref{fig:resolve}b show the path that would be taken if the
system passed through the two-phase region entirely with a common
stress in the two phases.  This scenario follows from minimizing the
chemical potential, and its correctness, of course, should be further
examined by the full time evolution of the original dynamic equations.

It is probable that boundary conditions also play a role. Consider a
Couette device.  Typically the walls provide uniform boundary
conditions in the azimuthal direction, while the slight inhomogeneity
of Couette flow induces an asymmetry between the inner and outer
cylinders.  The slightly higher stress near the inner wall provides a
preference for the high strain rate nematic phase, and hence might
enhance the stability of common stress phase separation.  Similarly,
the intrinsic inhomogeneity (although weaker) in cone-and-plate
rheometry induces a preference for the common stress interfacial
configuration \cite{BritCall97}.

We are also unable to say anything about the number or spacing of
bands.  Analogies with equilibrium systems suggest that phase
separation would coarsen until the system formed two bands at
different strain rates (for phase separation at a common stress).
This is the behavior seen in visualizations of flow in Couette,
cone-and-plate, and pipe geometries, where the intrinsic inhomogeneity
provides a ``seed'' for macroscopic phase separation
\cite{berret94b,Call+96,MairCall96,BritCall97,MairCall97,boltenhagen97a}.
Recent visualization of banding in lamellar surfactant systems
\cite{Bonn+98} indicate that phase separation at a common strain rate
can exhibit bands (disc-like bands in Couette flow) whose initial
spacing depends on the applied strain rate and coarsen in time. Unlike
common stress bands, which are expected to (and do) form macroscopic
bands in Couette flow, there is no boundary effect in Couette (aside
from perhaps sedimentation) which would encourage common strain rate
bands to coalesce readily. Normal stresses may play a role in this
process.

%%%%%%%%%%%%%%%%%%%%%%%%%%%%%%%%%%%%%%%%%%%%%%%%%%%%%%%%%%%%%%%
\subsection{Stability at prescribed stress or prescribed strain
  rate} \label{sec:spinodals}
%%%%%%%%%%%%%%%%%%%%%%%%%%%%%%%%%%%%%%%%%%%%%%%%%%%%%%%%%%%%%%%
In calculating the phase diagrams we have calculated the stability of
the fluid under conditions of either fixed strain rate or fixed
stress. These limits of stability, analogous to spinodals in
equilibrium systems, are displayed in Figures~\ref{fig:tie}
and~\ref{fig:ties}. Note that the stability limits and critical points
differ, depending on the control variable (stress $\sigma$ or strain
rate $\dot{\gamma}$). To see why this is, note that schematically the
dynamical equations of motion have the form
\begin{align}
  \partial_t \textbf{x} &= \textbf{f}(\textbf{x},\dot{\gamma}) \label{eq:dtx}\\
  \sigma &= g(\textbf{x},\dot{\gamma})\label{eq:dtsigma}
\end{align}
where $\textbf{x}$ comprises the dynamical variables (order parameter
$\boldsymbol{Q}$ and composition $\phi$). The second equation relates
the stress to strain rate and dynamical variables at steady state (or
in the zero Reynolds number limit), which implies that the strain rate
$\dot{\gamma}$, for a given stress, is a function $\dot{\gamma} =
\dot{\gamma}(\sigma,\textbf{x})$.  Consider fluctuations about a
steady state $\textbf{x}_0$: $\textbf{x} = \delta\textbf{x} +
\textbf{x}_0$. The dynamics for the fluctuation obeys
\begin{align}
  \partial_t\,\delta\textbf{x} &=
  \left\{\frac{\partial\textbf{f}}{\partial\textbf{x}} +
    \frac{\partial\textbf{f}}{\partial\dot{\gamma}}
    \frac{\partial{\dot{\gamma}}} {\partial\textbf{x}}
  \right\}\cdot\delta\textbf{x}\\
  &\equiv \left\{\textsf{M}_{\gamma} +
    \delta\textsf{M}_{\sigma}\right\}\cdot\delta\textbf{x} 
\end{align}

The limit of stability for common strain rate is calculated using the
fluctuation matrix $\textsf{M}_{\gamma}$, while the limit of stability
for common stress was calculated using $\textsf{M}_{\gamma} +
\delta\textsf{M}_{\sigma}$. These correspond to different stability
criteria.

The question of which spinodal could be observed in an experiment
relies on the accuracy of prescribed stress and prescribed strain rate
rheometers.  For a rheometer operating at a prescribed strain rate,
then if $\delta\textbf{x}$ goes unstable through $\textsf{M}_{\gamma}$
(in Eq.~\ref{eq:dtx}), the stress increases due to Eq.~(\ref{eq:dtsigma})
and no attempt is made to control it, leading to instability.
However, consider a rheometer which maintains a prescribed stress.  If
the system goes unstable in Eq.~(\ref{eq:dtx}), the bulk stress will
change due to Eq.~(\ref{eq:dtsigma}).  A sensitive and fast enough
rheometer will respond by adjusting the strain rate accordingly, to
maintain the imposed stress. Hence, instability would be determined by
the sum $\textsf{M}_{\gamma} + \delta\textsf{M}_{\sigma}$.
   
Similarly, in equilibrium systems a locus of stability may be defined
by, for example, the diverging of isothermal or adiabatic, or isobaric
or isochoric, response functions (or the vanishing of the appropriate
modulus).  For example, the isothermal and adiabatic compressibilities
$K_T$ and $K_S$ differ by a term proportional to the quotient of the
square of the thermal expansion coefficient $\alpha_p$ and the
isobaric heat capacity $c_p$:
\begin{equation}
  \label{eq:7}
  K_T - K_S = \frac{vT\alpha_p^2}{c_p},
\end{equation}
where $v$ is the specific volume. $K_T^{-1}$ vanishes along the
spinodal line $v_s(T)$, while it is evident that $K_S^{-1}$
(proportional to the sound speed) does not.  However, in equilibrium,
the critical point is uniquely defined in phase space, which is
related to the fact that, for example, pressure is a unique
\emph{function} of the volume, and is in fact a state variable.
Conversely, we can see from the shape of the stress--strain rate
curves for the Doi model (\emph{e.g.}  Fig.~\ref{fig:constit}c-e),
that the stress can be a multivalued function of strain rate;
\emph{i.e.} it is not a state function. Hence there is no compelling
reason to expect critical points at imposed strain rate to be the same
as critical points at imposed stress.  Similarly, the {\em true\/}
spinodal, or locus of stability, is uniquely defined in an equilibrium
system because of the convexity requirement on the entropy
\cite{debenedetti}, and there is no such universal convexity
requirement (barring entropy production, which is minimized only under
restricted conditions, and only locally rather than globally) for
non-equilibrium systems.

%%%%%%%%%%%%%%%%%%%%%%%%%%%%%%%%%%%%%%%%%%%%%%%%%%%%%%%%%%%%%%%%%%%%%%%%%%%%
\subsection{An analogy with equilibrium systems?}
%%%%%%%%%%%%%%%%%%%%%%%%%%%%%%%%%%%%%%%%%%%%%%%%%%%%%%%%%%%%%%%%%%%%%%%%%%%%
The liquid crystalline suspension under flow, indeed any system which
undergoes a macroscopic bulk flow-induced phase transition, is
analogous to an equilibrium ternary system comprising species $A$,
$B$, and solvent. In our case, the roles of $A$ and $B$ are played by
the rigid rod composition $\phi$ and either the stress $\sigma$ or
strain rate $\dot{\gamma}$, depending on the nature of the phase
separation. For phase separation at common stress, the phase diagram
in the stress-composition plane $\sigma-\phi$ is analogous to the
$\mu_A-\phi_B$ plane for the equilibrium system, while the
$\dot{\gamma}\!-\!\phi$ plane is analogous to the $\phi_A\!-\phi_B$
plane. In either case, the density variables,
$\left\{\dot{\gamma},\phi\right\}$ in flow and
$\left\{\phi_A,\phi_B\right\}$ in the analogous equilibrium system,
are different in the two coexisting phases. The slope of the
``plateau'' in the $\sigma\!-\dot{\gamma}$ plane, as the system
traverses the two-phase region of the phase diagram, is analogous to a
slope in the $\mu_A\!-\!\phi_B$ plane, the latter indicating that the
chemical potential (or osmotic pressure) of the two phases varies
across the coexistence region.

Can this analogy be extended to the possibility of phase separation at
common stress \emph{or} common strain rate? Certainly, one can
consider a ternary system under conditions of either imposed $\phi_A$
or imposed $\mu_A$, for which one generally expects difference
spinodal lines. That is, the spinodal is determined by the instability
of a matrix in the two-dimensional space spanned by $\phi_A$ and
$\phi_B$, and fixing $\phi_A$ or $\mu_A$ projects this instability onto
different subspaces.  Experimental conditions may dictate that the
spinodal line under fixed $\phi_A$ is more likely to be seen by, since
$\phi_A$ is conserved and cannot equilibrate quickly to satisfy an
imposed $\mu_A$. However, we are not aware of any ternary equilibrium
system for which the equilibrium coexistence conditions can differ;
that is, equilibrium is \emph{always} specified by equality of $\mu_A$
and $\mu_B$ in the two phases, and never by equality of $\phi_A$.
%%%%%%%%%%%%%%%%%%%%%%%%%%%%%%%%%%%%%%%%%%%%%%%%%%%%%%%%%%%%%%%%%%%%%%%%%%%%
\section{Summary}
%%%%%%%%%%%%%%%%%%%%%%%%%%%%%%%%%%%%%%%%%%%%%%%%%%%%%%%%%%%%%%%%%%%%%%%%%%%%

In this work we have proposed a straightforward phenomenological
extension to the Doi model for a solution of rigid rod particles. We
have added entropic terms, and included inhomogeneous terms in order
to calculate, for the first time, phase separation in shear flow. The
main results of this study are as follows:
\begin{enumerate}
\item Phase separation may occur under conditions of common stress
  \emph{or} common strain rate, with different interface orientations
  with respect to flow geometry for the two cases.
\item Although both phase separations are possible, the phase diagrams
  in the $\mu\!-\!\sigma_{xy}$ and $\mu\!-\!\dot{\gamma}$ planes
  (Fig.~\ref{fig:mu}) suggest that phase separation at a common strain
  rate is metastable. This can be traced, for this model, to the
  shear-thinning character of the transition. For a shear thickening
  transition an equivalent argument suggests that (if both are
  kinematic possibilities) common stress phase separation is
  metastable with respect to strain rate phase separation.
\item The limits of stability (``spinodals'') and critical points for
  systems at prescribed stress and prescribed strain rate differ; the
  difference of spinodals is similar to equilibrium behavior, while
  the difference of critical points is related to the fact that
  neither stress or strain rate are always unique state functions.
\item An argument based on minimizing the chemical potential predicts
  a complex crossover from common stress to common strain rate phase
  separation for controlled strain rate experiments.  The veracity of
  this assumption is unknown.
\item We have calculated phase coexistence among three phases
  (paranematic \textsf{I}, flow-aligning nematic \textsf{N}, and
  log-rolling nematic \textsf{L}), where only two phases existed in
  equilibrium. We expect \textsf{I-N} phase coexistence to be the
  stable configuration (Fig.~\ref{fig:tieboth}), although \textsf{I-L}
  phase coexistence could exist as a metastable state with approprate
  preparation. We do not expect three phase coexistence for this
  model.
\item We have demonstrated how to calculate the mean
  stress--strain-rate relationship $\bar{\sigma}(\bar{\dot{\gamma}})$
  in the coexistence region. The shape of
  $\bar{\sigma}(\bar{\dot{\gamma}})$ is determined by the composition
  and strain rates of the coexisting phases \cite{schmitt95}.
\item A phase-separated system can exhibit an apparently unstable
  constitutive relation, with negative slope
  $\partial\sigma_{xy}/\partial\dot{\gamma}$. Experiments have
  accessed such negative slope composite curves under controlled
  stress (rather than controlled strain rate) conditions \cite{HBP98}.
\item Our method of solution is general and relies on the existence of
  a set of dynamical equations of motion for the structural order
  parameter of the particular transition, including the dynamic
  response to inhomogeneities.
\item For $\lambda=g/K$ finite the phase boundaries we have found are,
  within our accuracy, independent of the relative magnitude of the
  gradient terms in our free energy. Although this suggests that, for
  the restricted set of inhomogeneities we have incorporated, a
  selection criterion exists involving only the homogeneous equations
  of motion, this is not true in general for complex fluids in flow
  \cite{jsplanar}. For $\lambda=\infty$ the phase boundaries are
  slightly shifted.
\item Studies at different aspect ratios suggest that shear flow
  enhances polydispersity effects relative to their effect on
  equilibrium phase boundaries.
\end{enumerate}

We close by enumerating several open questions. First, systems such as
wormlike micelles probably possess some combination of a perturbed
isotropic-nematic transition and a dynamic instability of the
molecular constitutive relation. It is conceivable that suitable
compositions of these systems could yield a
stress--strain-rate--composition surface (Fig.~\ref{fig:3D}) with
multiple folds. Second, it would be desirable to have a model
shear-thickening system in which to calculate properties of banded
flows, to compare and contrast with the shear-thinning system studied
here and to understand experiments on a wide range of systems,
including clays and surfactant systems.  Third, we have not addressed
the number and possible coarsening of bands and band configurations;
and the kinetics of phase separation has hardly been treated
theoretically \cite{berret94b}, with experiments also at an early
stage \cite{berret94b,grand97,Berr97}. Finally, we do not yet know the
conditions which may, if at all, distinguish between common stress or
common strain rate phase coexistence.

%%%%%%%%%%%%%%%%%%%%%%%%%%%%%%%%%%%%%%%%%%%%%%%%%%%%%%%%%%%%%%%%%%%%%%%%%%%%%
\acknowledgements
%%%%%%%%%%%%%%%%%%%%%%%%%%%%%%%%%%%%%%%%%%%%%%%%%%%%%%%%%%%%%%%%%%%%%%%%%%%%%
It is a pleasure to acknowledge helpful conversations and
correspondence with R. Ball, J.-F. Berret, G. Bishko, D. Bonn, M.
Cates, F. Greco, J. Harden, S.  Keller, G. Leal, T. McLeish, D. Pine,
G. Porte, O. Radulescu, N. Spenley, L. Walker, and X.-F.  Yuan.  CYDL
acknowledges funding from St. Catharine's College, Cambridge and the
(Taiwan) National Science Council (NSC 88-2112-M-008-005).
%%%%%%%%%%%%%%%%%%%%%%%%%%%%%%%%%%%%%%%%%%%%%%%%%%%%%%%%%%%%%%%%%%%%%%%%%%%%%
%%%%%%%%%%%%%%%%%%%%%%%%%%%%%%%%%%%%%%%%%%%%%%%%%%%%%%%%%%%%%%%%%%%%%%%%%%%%%
% \bibliography{misc,books,articles,many_copolymer,worms,capp,callaghan,schoot,zubarev,worms2,onions,newbib}
% \bibliographystyle{prsty}
%%%%%%%%%%%%%%%%%%%%%%%%%%%%%%%%%%%%%%%%%%%%%%%%%%%%%%%%%%%%%%%%%%%%

%%%%%%%%%%%%%%%%%%%%%%%%%%%%%%%%%%%%%%%%%%%%%%%%%%%%%%%%%%%%%%%%%%%%

%%%%%%%%%%%%%%%%%%%%%%%%%%%%%%%%%%%%%%%%%%%%%%%%%%%%%%%%%%%%%%%%%%%%
\end{multicols}
\end{document}